\documentclass[a4paper,fleqn,usenatbib]{mnras}

\usepackage[T1]{fontenc}
\usepackage{ae,aecompl}

\usepackage[normalem]{ulem}

\usepackage{graphicx} % Including figure files
\usepackage{amsmath}    % Advanced maths commands
\usepackage{amssymb}  % Extra maths symbols
\usepackage{tikz}
\usepackage{caption}
\usepackage{subcaption}
\usepackage{relsize}
\usepackage{booktabs}
\usepackage{newtxtext,newtxmath}
\title[The radio outflow in NGC\,2329]{The peculiar WAT NGC\,2329 with Seyfert/FR~I-like radio lobes}
\author[S. Das et al.]{S. Das$^{1}$\thanks{\href{mailto:soumyadeep.das.phy14@iitbhu.ac.in}{soumyadeep.das.phy14@iitbhu.ac.in} (SD)},
P. Kharb$^{2}$\thanks{\href{mailto:kharb@ncra.tifr.res.in}{kharb@ncra.tifr.res.in} (PK)},
R. Morganti$^{3,4}$\thanks{\href{mailto:morganti@astron.nl}{morganti@astron.nl} (RM) }, 
S. Nandi$^{2}$\thanks{\href{mailto:snandi@ncra.tifr.res.in}{snandi@ncra.tifr.res.in} (SN) }\\
$^{1}$Indian Institute of Technology (BHU) Varanasi, Varanasi 221005, Uttar Pradesh, India\\
$^{2}$National Centre for Radio Astrophysics - Tata Institute of Fundamental Research, Pune University Campus, Post Bag 3, Ganeshkhind, Pune 411007, India\\
$^{3}$Kapteyn Astronomical Institute, University of Groningen, Landleven 12, NL-9747 AD Groningen, The Netherlands \\
$^{4}$ASTRON, the Netherlands Institute for Radio Astronomy, Postbus 2, NL-7990 AA Dwingeloo, The Netherlands}

\pubyear{2021}
\begin{document}
\label{firstpage}
\pagerange{\pageref{firstpage}--\pageref{lastpage}}
\maketitle

\begin{abstract}
We report the complex radio properties of the radio galaxy NGC\,2329 that resides in the centre of the galaxy cluster Abell\,569. For this study, we have used archival data from the Very Large Array (VLA) at various resolutions and frequencies, as well as the Very Long Baseline Array (VLBA). While the wide-angle tailed (WAT) Fanaroff-Riley type I (FR~I) radio morphology of the source has been discussed widely in the literature, the nature of the inner lobes has not been as widely discussed. In particular, we note that the inner lobes resemble the bubble-like radio structures observed in Seyfert galaxies. Polarization-sensitive data from the VLA clearly show magnetic field structures consistent with FR~Is for the outer lobes and Seyferts for the inner lobes in NGC\,2329. FR~Is are classified as radio-loud (RL) active galactic nuclei (AGN) and Seyferts as radio-quiet (RQ) AGN, making this source unique. The VLBA shows a one-sided radio jet suggesting a relativistic pc-scale outflow, leading into the inner lobes. Electron lifetime estimates suggest that the outer FR~I-like lobes are nearly twice as old ($\sim$45~Myr) as the inner Seyfert-like lobes ($\sim$25~Myr). Gas inflow in this merging cluster seems to have rejuvenated the AGN about $\sim$25~Myr ago, and may have caused a change in the accretion disc state. The complex composite radio morphology of NGC\,2329 suggests that the RL/RQ dichotomy is a function of time in the lifecycle of an AGN.
\end{abstract}


\section{Introduction} \label{section:intro}    
Active galactic nuclei (AGN) are highly compact sources of broad-band radiation characterized by extremely high luminosities (upto $\mathrm{L_{bol}}\approx10^{48}$~{erg}~s$^{-1}$). They are powered by mass accretion on to supermassive black holes \citep[SMBH;][]{Rees1984}. The radio loudness parameter $(R)$ of an AGN is defined as the ratio of the 5-GHz radio flux density to the optical flux density in the 4400~\AA~{\it B}-Band \citep{Kellermann1989}. AGN with $R>10$ are categorized as radio-loud (RL), while those with $R\le10$ fall into radio-quiet (RQ) class {\citep[see however][for a recent review on this topic]{Padovani2017}.} Radio emission in RL AGN is predominantly from non-thermal synchrotron emission mechanism in jets and lobes, while in RQ AGN like Seyfert galaxies and Low-Ionization Nuclear Emission-Line Regions \citep[LINERs;][]{Heckman1980}, {thermal emission from stellar activity in the host galaxy may also contribute} to the radio emission, along with non-thermal synchrotron emission from supernovae, AGN jets and winds \citep{Gallimore2006,Berton2020,Sebastian2020,Silpa2020}. While RL AGN outflows can span Mpc and intergalactic scales, the radio outflows in RQ AGN typically range from 10 to 100 pc to tens of kpc scales \citep{Baum1993,Morganti1999,Gallimore2006,Baldi2018,Jarvis2019}.

The active phase of an RL AGN can typically last $\sim10^{7}-10^{8}$~yr \citetext{see review by \citealp{Morganti2017} and references therein; \citealp{Haehnelt1993,Parma1999,Parma2007}} and is characterized by the presence of a radio core, jets and extended lobes \citep{Urry1995,Shulevski2015}. There is considerable evidence in the literature about AGN activity being episodic \citep[e.g.][]{Schoenmakers2000,Kharb2006,Nandi2019,Jurlin2020,Sebastian2020}. {During the dormant phase of an AGN, core emission may disappear and lobes} without means of replenishment lose energy due to synchrotron radiation, inverse Compton (IC), and adiabatic losses \citep{vanderLaan1969,Shulevski2015,Brienza2017,Godfrey2017,Hardcastle2018}. The lobes start to age into a `remnant' structure, characterized by the steepening of the radio spectrum. Re-activation of the AGN could produce outflows at a different orientation from before, depending on the orientation of the newly formed accretion disc \citep[e.g.][]{Sikora2007}, leading to new jets/lobes, coexisting with the larger remnant emission \citep[e.g.][]{Roettiger1994,Kharb2016,Brienza2020,Jurlin2020}. Recent work by \citet{Bajraszewska2020,Nyland2020,Wolowska2021} have highlighted the fact that the radio-loudness itself may be a function of the epoch when the AGN is observed, blurring the RL/RQ divide.

Here, we present results for the peculiar radio galaxy NGC\,2329, which shows signatures of episodic AGN activity. The first activity episode produces a radio outflow resembling an RL AGN \citep[viz., a Fanaroff-Riley type I (FR~I) radio galaxy;][]{Fanaroff1974} while the second activity episode produces an outflow resembling a RQ AGN (viz., a Seyfert galaxy). This second outflow makes the `core' of the larger FR~I outflow. Both these outflows are clearly delineated in intensity. We have made use of archival Karl G. Jansky Very Large Array (VLA) and Very Long Baseline Array (VLBA) data to look at the source at different frequencies and spatial scales. This is the first attempt at distinguishing the various radio structures in total intensity and spectral indices in this peculiar galaxy. We discuss what our results imply for the larger question of the RL-RQ divide in AGN and the episodic nature of the AGN in NGC\,2329.

This paper is organized in the following way. Section \ref{section:ngc2329} summarizes the past work on the source. Section \ref{section:datareduction} details the data reduction and imaging procedures used in this study. Section \ref{section:results} evaluates the images and corresponding results, followed by a discussion in Section \ref{section:discussion}. A summary of the study and concluding remarks follow in Section \ref{section:conclusion}. In the text, we have referred to the outer large-scale outflows as the FR~I structure, while we allude to the arcsecond scale inner lobes as the Seyfert source. We have used the cosmology with H$_{0}$ = 73 km~s$^{-1}$~Mpc$^{-1}$, $\Omega_\mathrm{M} = 0.27$, and $\Omega_\mathrm{vac} = 0.73$. At the redshift $z = 0.0196$ for NGC\,2329 \citep{Smith2000}, 1 arcsec corresponds to 0.381 kpc. The spectral index $\alpha$ is defined such that $S_\nu \propto \nu^\alpha$. 

\section{NGC\,2329} \label{section:ngc2329}
NGC\,2329 (also known as UGC\,3695) is the brightest cluster galaxy (BCG) of the nearby galaxy cluster Abell\,569 \citep{Peterson1970, Fanti1982, Wrobel1984}. A\,569 is likely a part of an extended Perseus supercluster filament \citep{Balkowski1989, Chamaraux1990} and is comprised of two major cluster condensations \citep{Fanti1982}. NGC\,2329 has been identified as the BCG of the Southern condensation. The dynamical state of A\,569 suggests that the cluster is not relaxed and is in an intermediate stage of evolution {\citep{Feretti1985}}. X-ray emission from A\,569 is dominated by emission from galaxies and point sources \citep{Abram1983,Sakelliou2000}. The distribution of the ICM is unclear. \cite{Abram1983,Feretti1985,Feretti1990} found hints of weak diffuse emission and an extended gaseous halo {centred around the brightest cluster galaxy}. The observed weak diffuse emission is similar to that seen in merging clusters \citep{Rossetti2011,Drabent2015,VanWeeren2019}.

No cooling core has been observed in this source \citep{Liuzzo2010}. 
{The host galaxy has been classified as a lenticular (S0) \citep[RC3;][]{DeVaucouleurs1991}. While the radial luminosity profile of the galaxy satisfies the $r^{1/4}$ power law \citep{Schombert1986} similar to bulge-dominated galaxies, {\it Hubble Space Telescope (HST)} images \citep{Kleijn1999} reveal a small inclined dust disc in the centre consistent with its S0 galaxy classification.} \cite{Kleijn1999} found that the nucleus of NGC\,2329 was bluer than the surroundings. The galaxy had exceptional blue $V-I$ color as compared to the other giant ellipticals in their sample. \cite{Kleijn1999,Kleijn2002} detected a small $\sim$0.3~arcsec (114~pc) optical jet-like extension emerging from the core at position angle (PA) $322\degr$.

A list of physical parameters pertaining to NGC\,2329 is presented in Table \ref{table:physicalparam}. VLA observations by \citet{Wrobel1984} found a compact core at the centre with symmetric weak diffuse double-sided outflows with a total linear extent of 40~arcsec ($\sim$15~kpc) extending away from it. VLA observations of the source at 1.4 and 4.9~GHz were presented by \citet[][see their fig.~8]{Kharb2014}, who noted the uniqueness of this source in their FR~I galaxy sample due to its lowest radio loudness parameter ($R$), which was overlapping with the $R$ values of Seyfert galaxies. They also noted its total lobe extent matching those of Seyfert lobes (typically in the range of $\sim$10~kpc).

\begin{table}
\caption{NGC\,2329 : physical parameters}
\begin{center}
\begin{tabular}{lrr}\hline
Parameter & Value & Ref. \\  \hline
${M_\mathrm{BH} (\times 10^8~\mathrm{M}_{\sun}}$) &  4.92 & 1 \\
Absolute {\it B}-band magnitude              & 13.6~~               & 2 \\
$\mathrm{H}\alpha$ luminosity ($\times 10^{39}\mathrm{erg~s}^{-1}$)&   6.0~~ & 3 \\
Star formation rate ($\times 10^{-2}~\mathrm{M_{\sun}~yr^{-1}}$) \hspace{1.2cm} & 4.11 & \hspace{0.7cm} 4\\
\hline
\end{tabular}
\end{center}

{\it References}. (1) \citet{NoelStorr2007}, (2) \citet{Kharb2014}, (3) \citet{Kleijn1999}, (4) \citet{Vaddi2016}.
\label{table:physicalparam}
\end{table}

WSRT observations by \cite{Feretti1985} showed a large-scale wide-angle tail (WAT) structure with two diffuse lobes of uniform surface brightness, each extending about 2.5 ($\sim$57~kpc) to 2.9~arcmin ($\sim$67~kpc) away from the core. They also observed a region of very low surface brightness between the lobes at 0.6~GHz. The source displayed one-sided core-jet VLBI morphology \citep{Xu2000,Kharb2012}. Einstein Observatory found that the X-ray emission in A\,569 is centred around the BCG (NGC\,2329), with weak evidence for diffuse emission and an extended gaseous halo, possibly accreting onto the system \citep{Abram1983,Feretti1985}. \cite{Fanti1982,Feretti1985} have classified the host as a D or cD galaxy. 

The WAT morphology in NGC\,2329 is curious in itself. \citet{Fanti1982} calculated the total velocity dispersion of A\,569 as 5715~km~s$^{-1}$, and also the radial velocity of NGC\,2329 as 5570~km~s$^{-1}$. Thus, the velocity of the galaxy across the cluster is very low ($\Delta v$ = 145~km~s$^{-1}$), effectively ruling out the possibility of ram pressure being the reason behind the WAT structure \citep[eg.][]{Eilek1984,Burns1986,Sakelliou2000}. The possible causes of the WAT structure may be the large-scale motion of the ICM with respect to the galaxy due to either dynamic evolution of the cluster \citep[e.g.][]{Sparke1983}, or cluster collision, possibly between the two condensations of A\,569 \citep[e.g.][]{Patnaik1984,Sakelliou1999,Lakhchaura2011}.

\begin{table*}
\caption{Observation and imaging parameters for archival VLA and VLBA data, and flux densities and rms noise values estimated from images.}
\centering
\begin{tabular}{lcccrcrrc}\hline
\multicolumn{5}{c}{Observation parameters} & \multicolumn{2}{c}{Image parameters} & \multicolumn{2}{c}{Flux densities}\\
Telescope & \multicolumn{1}{c}{$\nu$} & \multicolumn{1}{c}{Project} & \multicolumn{1}{c}{Date} & \multicolumn{1}{c}{Exp.} & \multicolumn{1}{c}{Beam size} & \multicolumn{1}{c}{Beam PA}  & \multicolumn{1}{c}{Component flux} &  \multicolumn{1}{c}{rms}  \\ 
          &       & \multicolumn{1}{c}{code}    &      &      &               &            & \multicolumn{1}{c}{density}        &  \multicolumn{1}{c}{noise}   \\
 & \multicolumn{1}{c}{(GHz)} & & & \multicolumn{1}{c}{(min.)} &  & \multicolumn{1}{c}{$(\overset{\circ}{ })$} & \multicolumn{1}{c}{(mJy)} & \multicolumn{1}{c}{(mJy beam$^{-1}$)} \\ 
{(a)} & \multicolumn{1}{c}{(b)} & \multicolumn{1}{c}{(c)} & \multicolumn{1}{c}{(d)} & \multicolumn{1}{c}{(e)} & \multicolumn{1}{c}{(f)} & \multicolumn{1}{c}{(g)} & \multicolumn{1}{c}{(h)} & \multicolumn{1}{c}{(i)}  \\ \hline
VLA:A   &  1.43 & AB0920 & 1999 Jul 18 & 30~~~~ & $1.39 \times  1.22$ arcsec$^2$ &  24.2~~  & Core              ~~:~~165 &  0.04 \\
        &      &        &              &    &                                      &        & N. Seyfert Lobe   ~~:~~~ 63 &       \\
        &      &        &              &    &                                      &        & S. Seyfert Lobe   ~~:~~~ 43 &       \\
VLA:B   &  1.51 & AC0445 & 1995 Dec 22 &  5~~~~ & $4.30 \times 4.13$ arcsec$^2$ & $-76.7$~~   & Core            ~~:~~275  &  0.08 \\
        &      &        &              &    &                                      &        & N.W. FR~I Lobe    ~~:~~~ 59  &       \\
        &      &        &              &    &                                      &        & S. FR~I Lobe      ~~:~~~ 86  &       \\
VLA:C   &  1.49 & AB0412 & 1986 Dec 5 &  3~~~~ & $16.62 \times  13.55$ arcsec$^2$ & 60.1~~   & Core             ~~:~~334  &  0.28 \\
        &      &        &              &    &                                      &        & N.W. FR~I Lobe     ~~:~~198  &       \\
        &      &        &              &    &                                      &        & S. FR~I Lobe       ~~:~~241  &       \\
VLA:B$+$C$^*$ &  1.49 & ...   & ...         & ...~~~~ & $16.74 \times 13.61$ arcsec$^2$ & 60.8~~   & Core              ~~:~~336  &  0.23 \\
        &      &        &              &    &                                      &        & N.W. FR~I Lobe     ~~:~~201  &       \\
        &      &        &              &    &                                      &        & S. FR~I Lobe       ~~:~~244  &       \\
VLA:B   &  4.71 & AC0445 & 1995 Dec 22  &  4~~~~ & $1.68 \times 1.32$ arcsec$^2$ &  80.7~~    & Core             ~~:~~114  &  0.04 \\
        &      &        &              &    &                                      &        & N. Seyfert Lobe  ~~:~~~ 29  &       \\
        &      &        &              &    &                                      &        & S. Seyfert Lobe  ~~:~~~ 19  &       \\
VLA:D   &  4.71 & AC0557 & 2000 Oct 1 &  4~~~~ & $15.66 \times 12.62$ arcsec$^2$ & 41.8~~   & Core             ~~:~~209  &  0.08 \\
        &      &        &              &    &                                      &        & N.W. FR~I Lobe    ~~:~~~ 54  &       \\
        &      &        &              &    &                                      &        & S. FR~I Lobe      ~~:~~~ 81  &       \\
VLA:AB  & 14.94 & AS0451 & 1991 Oct 13 &  4~~~~ & $0.34 \times 0.21$ arcsec$^2$ &  83.2~~    & Core $+$ Jet    ~~:~ 86$^a$, 93$^b$  & 0.09 
\vspace{0.2cm} \\ 
VLBA    &  2.27 & BS0103 & 2002 Jan 2  & 49~~~~ & 7.7 mas $\times$ 2.5 mas             &  $-12.9$~~     & Core $+$ Jet ~~:~ 50$^a$, 62$^b$  & 0.08 \\
VLBA    &  4.99 & BO0015  & 2003 Feb 14 & 28~~~~ & 2.5 mas $\times$ 1.4 mas & 12.7~~     & \hspace{0.6cm} Core $+$ Jet  ~~:~ 53$^a$, 72$^b$  & 0.08 \\ \hline
\end{tabular}
\begin{flushleft}
{Columns: (a) Telescope used, with array configuration for VLA datasets (b)~observing frequency in GHz; (c) VLA/VLBA project identification; (d) date of observation; (e) total exposure time summed over all scans in minutes; (f)~full width at half maximum (FWHM) of synthesized beam; (g) synthesized beam PA in degrees; (h) flux density of individual resolved components in mJy using {\tt TVSTAT} in \textsc{aips}, `Core' refers to the compact component visible in the respective images, NW. = north-west, N = north, S = south; (i) rms noise in image in mJy beam$^{-1}$. $^a$core peak intensity in mJy~beam$^{-1}$ using {\tt TVMAXF} in \textsc{aips}. $^b$Total flux density in Core + Jet in mJy using {\tt TVSTAT} in \textsc{aips}. $^*$VLA:B$+$C data set was created by combining 1.5-GHz VLA:B- and C- array data sets in \textsc{aips} (see Section \ref{section:outerlobes} for details). Project codes, dates, and exposure times of VLA:B- and C- data sets can be found in previous rows}
\end{flushleft}
\label{table:vla}
\end{table*}

\section{Radio Data Processing}\label{section:datareduction}
We have used data available in the historical VLA and VLBA archives {for a complete analysis of the various structures observed in NGC\,2329. VLA data presented here come from several array configurations, and to the best of our knowledge, are previously unpublished.\footnote{We have used the VLA data from Project AB0920 previously published in \citet{Kharb2014} to make new spectral index images utilizing additional radio data in this paper.}} The archival data were acquired at {1.5, 5, and 15~GHz} between 1985 and 2000. {We have carried out polarization calibration for the VLA B-array archival data set AC0445 from 1995 December 22, which had a good parallactic angle coverage for the primary flux density and polarization calibrator, 3C~286.} {Phase-referenced VLBA observations were acquired at 2.3 and 5~GHz.} Details of these observations are listed in Table \ref{table:vla}. 

VLA data processing was carried out by utilizing the standard calibration and imaging routines in the Astronomical Image Processing System \citep[\textsc{aips};][]{Greisen2003}. Several rounds of phase-only and phase$+$amplitude self-calibration were performed to remove any residual errors. Image parameters have been detailed in Table \ref{table:vla}. The VLBA data processing made use of the {\tt VLBARUN} routine with suitable parameters in \textsc{aips}, followed by a couple of rounds of phase-only and phase$+$amplitude self-calibration. Details of the image parameters are listed in Table~\ref{table:vla}. Finally, spectral index images of different scale VLA structures were obtained using the \textsc{aips} task {\tt COMB}. Images at the different frequencies were made at identical resolutions after matching the {\tt uv} spacings and made to be positionally coincident before creating the spectral index images. Pixels with intensity values below $3\sigma$ were blanked in {\tt COMB}. 

We used the \textsc{aips} task {\tt PCAL} to solve for the antenna `leakage' terms (D-terms) as well as source (i.e. 3C~286) polarization for the VLA B-array data set AC0445. The D-terms were typically a few percent at both 1.5 and 5~GHz. 3C~286 was also used to calibrate the polarization angle. The polarization intensity {\tt (PPOL)} and polarization angle {\tt (PANG)} images were created using Stokes `Q' and `U' images in the task {\tt COMB}. Pixels with intensity values below $3\sigma$ and angle errors $>10\degr$ were blanked while making {\tt PPOL} and {\tt PANG} images, respectively. We also created fractional polarization (FPOL) images using the polarized intensity and total intensity images in {\tt COMB} where pixels with $>10\%$ errors in FPOL were blanked. Noise images were created for {\tt PPOL}, {\tt PANG} and FPOL using task {\tt COMB}. Flux density, spectral index, and fractional polarization values reported in this paper have been obtained using \textsc{aips} verbs {\tt TVWIN+IMSTAT} or {\tt TVSTAT} on the appropriate images. A discussion of the image attributes follows in Section \ref{section:results}.

\begin{figure*}
\centering{
\includegraphics[width=8.3cm]{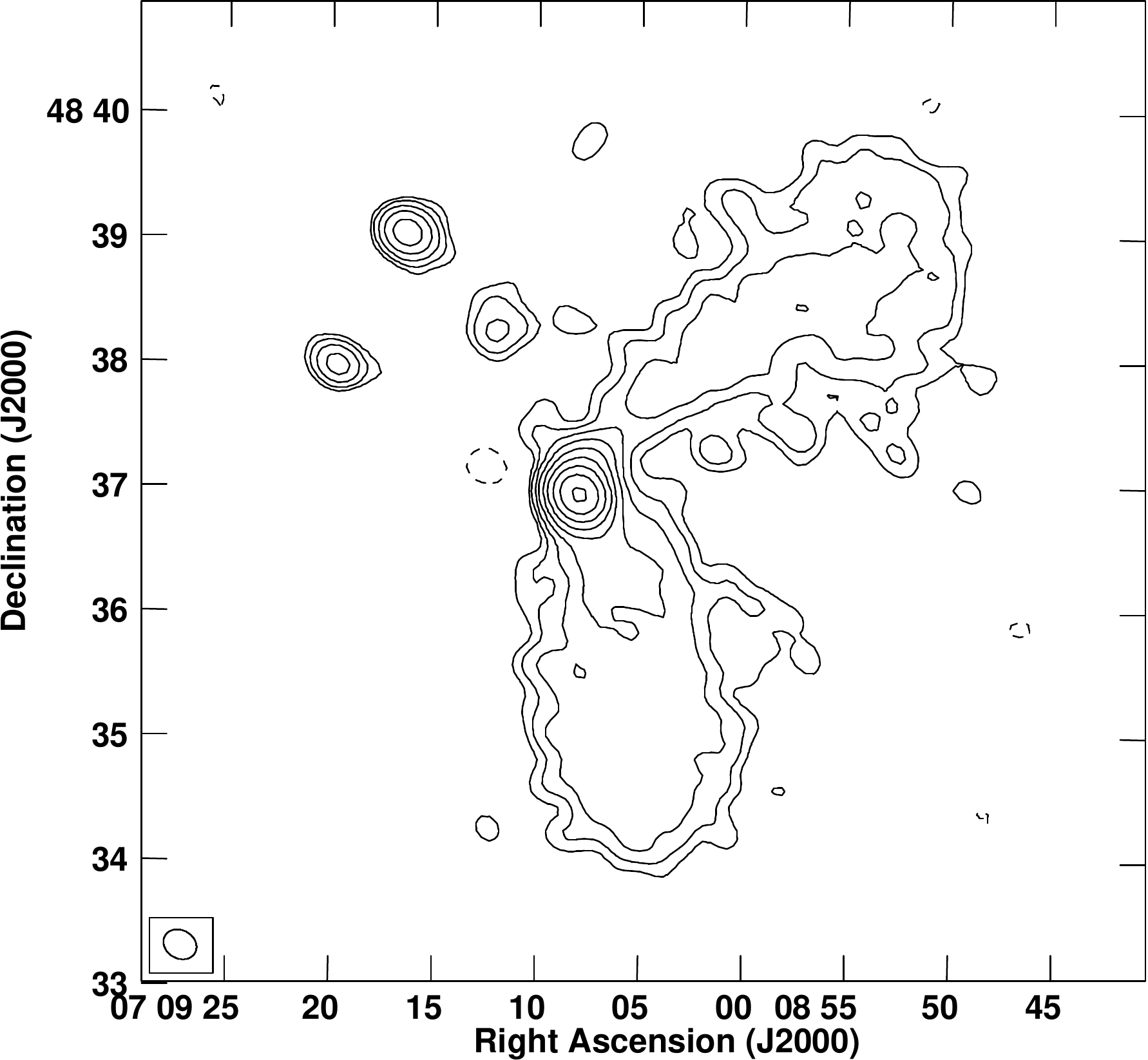}
\includegraphics[width=8.3cm]{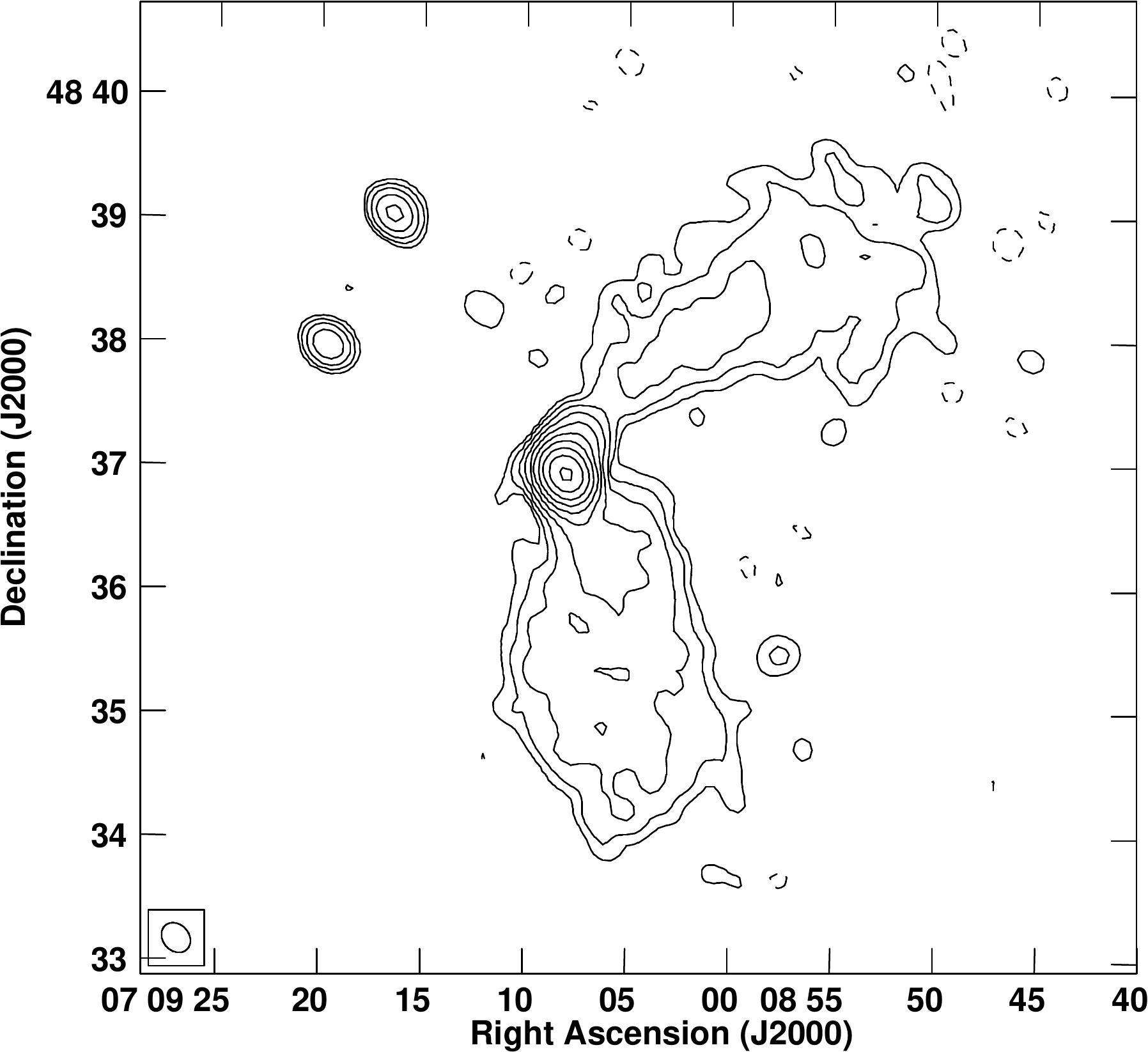}
\vspace{0.5cm}

\includegraphics[width=8cm]{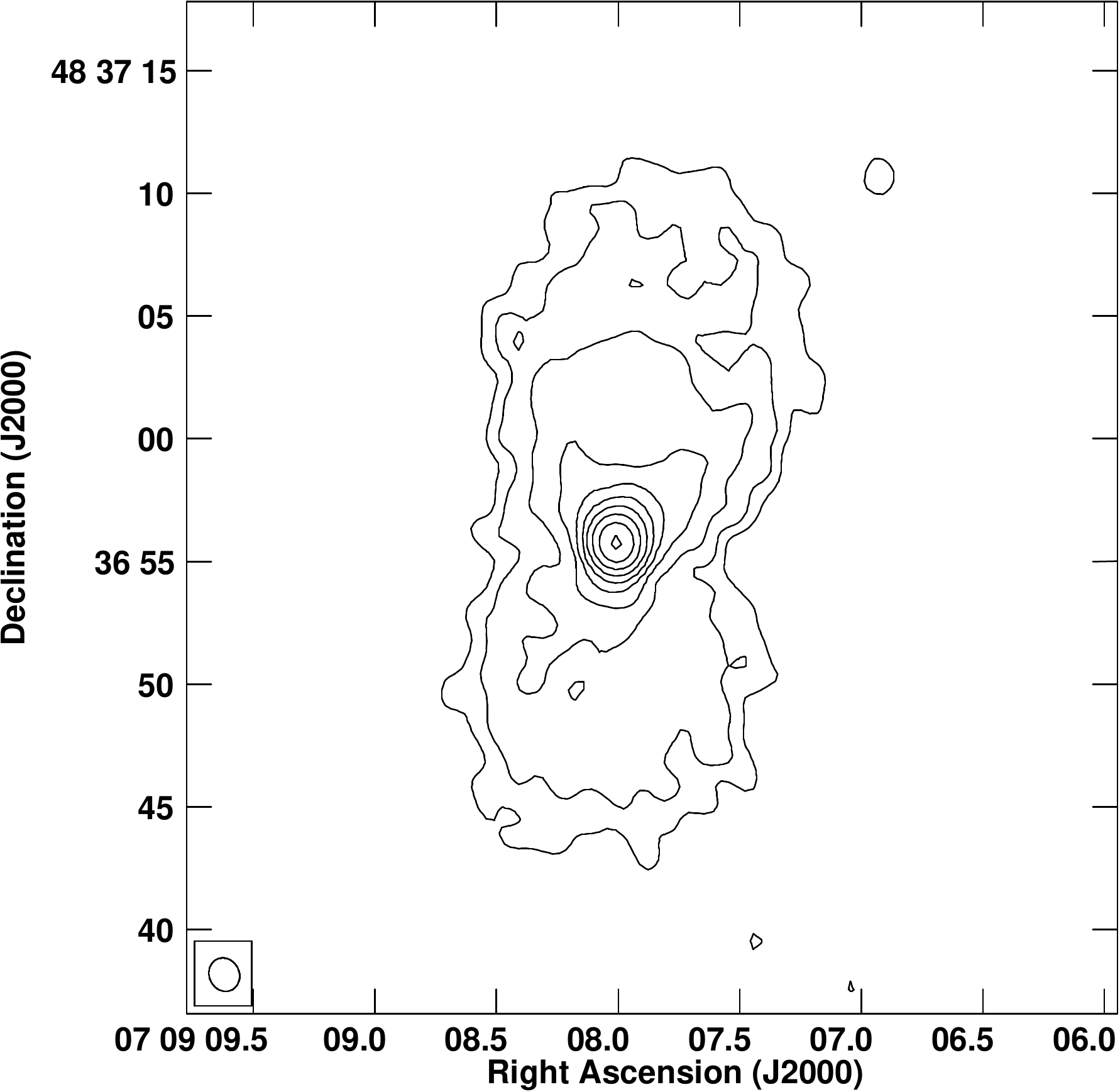}
\includegraphics[width=8.9cm]{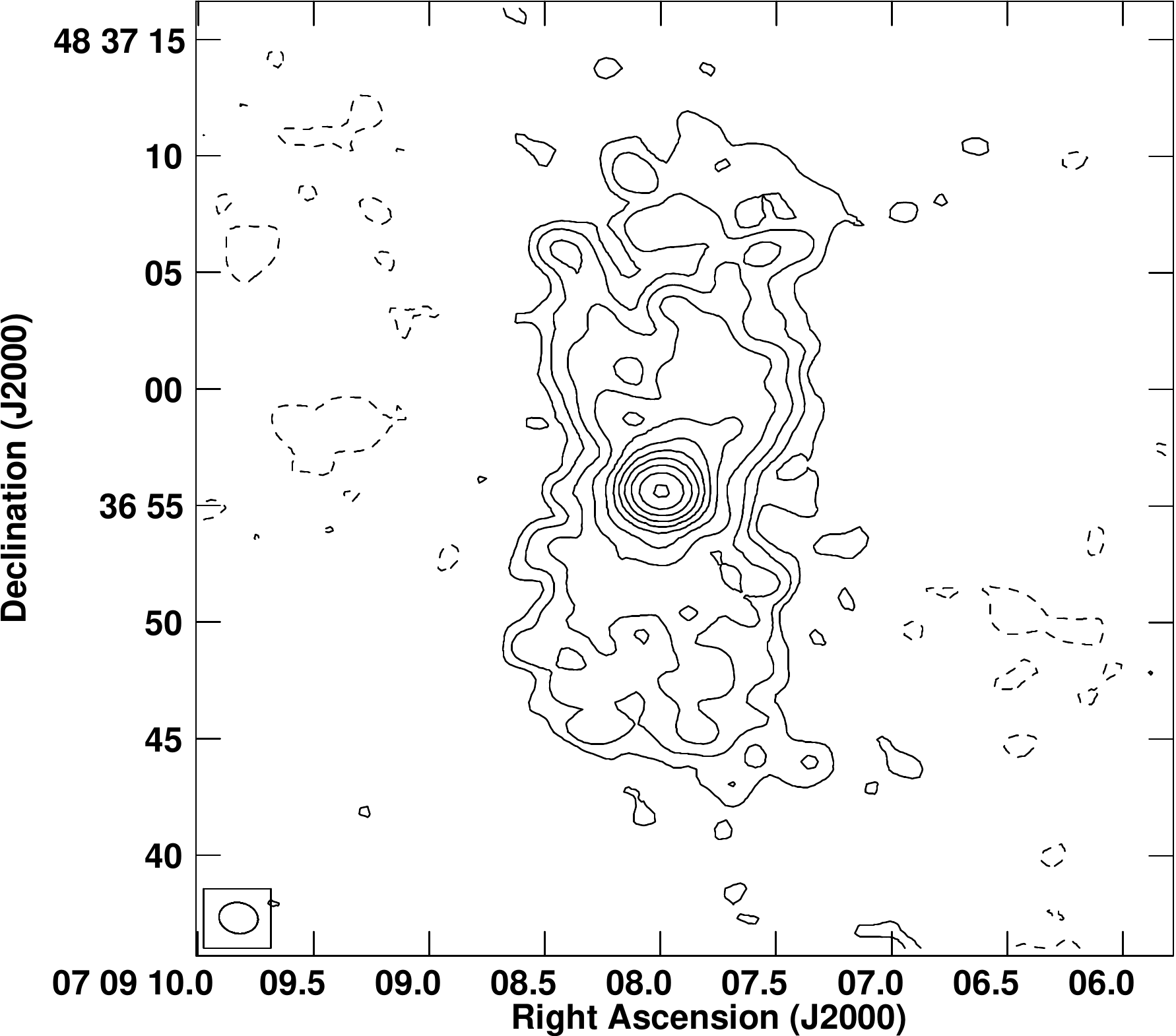}}
\caption{\small VLA contour images of NGC\,2329. Contours are in percentage of peak surface brightness and increase in steps of two. 
Top left-hand panel: {1.5-GHz} VLA C-array configuration image with lowest contour level and peak surface brightness, $\pm$0.35 per cent and 232~mJy~beam$^{-1}$, respectively. Beam size is $16.62 \times13.55$ arcsec$^2$ at PA $60\overset{\circ}{.}1$. 
Top right-hand panel: {5-GHz} VLA D-array image with the lowest contour level and peak surface brightness, $\pm$0.18 per cent and 160~mJy~beam$^{-1}$, respectively. Beam size is $15.66 \times 12.62$ arcsec$^2$ at PA $41\overset{\circ}{.}8$.
Bottom left-hand panel: {1.5-GHz} VLA A-array image with lowest contour level and peak surface brightness, $\pm$0.17 per cent and 130\,mJy beam$^{-1}$, respectively. Beam size is $1.39 \times 1.22$ arcsec$^2$ at PA $24\overset{\circ}{.}2$. 
Bottom right-hand panel: {5-GHz} VLA B-array image with lowest contour level and peak surface brightness, $\pm$0.09 per cent and 96~mJy~beam$^{-1}$, respectively. Beam size is $1.68 \times1.32$ arcsec$^2$ at PA $80\overset{\circ}{.}7$. 
}
\label{fig:vla-stokesi-collage}
\end{figure*}

\section{Results}\label{section:results}
\begin{figure}
\centering{
\includegraphics[width=8cm]{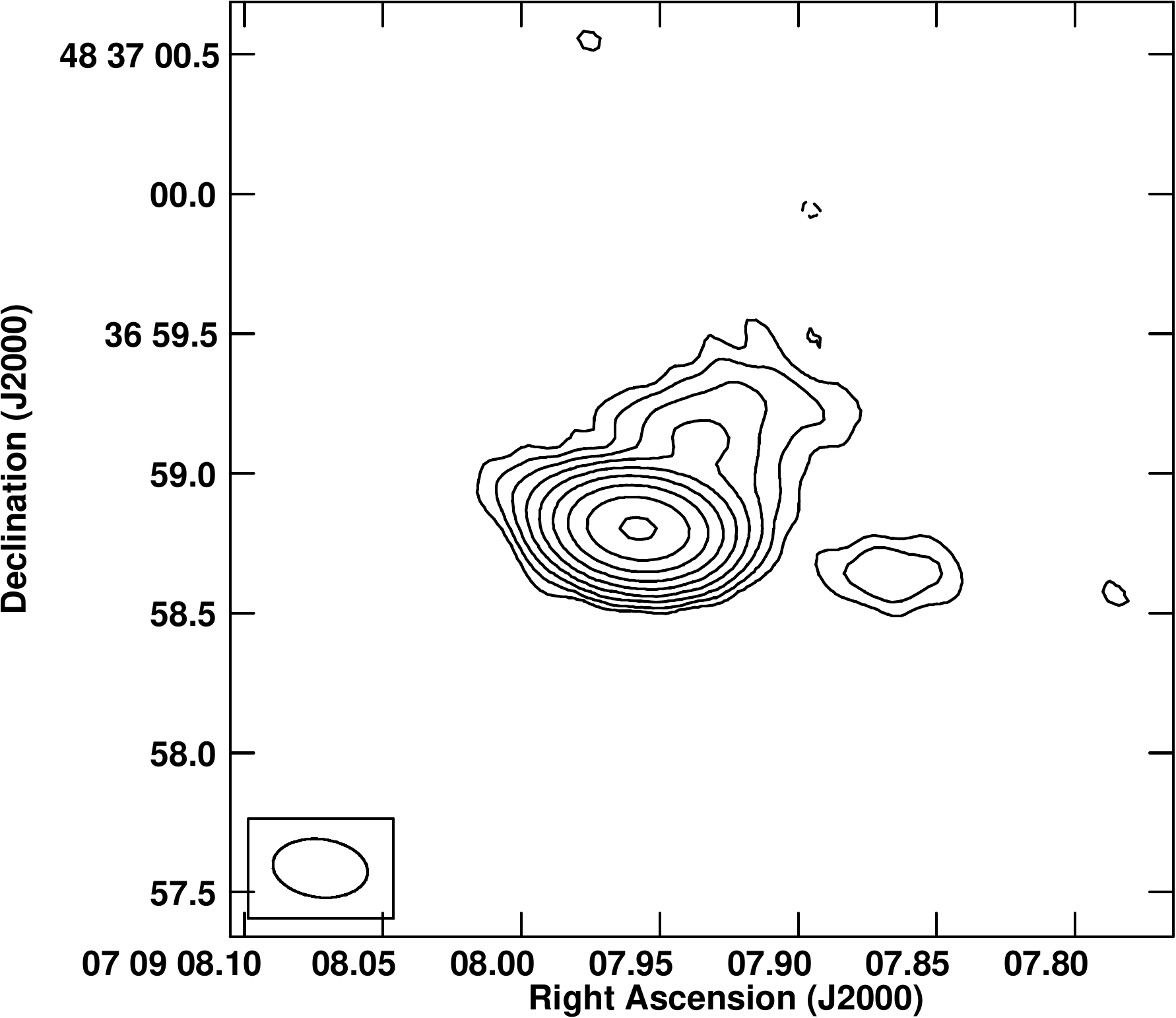}
\caption{15-GHz VLA AB-array contour image of NGC\,2329. Contours are in percentage of peak surface brightness and increase in steps of two. Lowest contour level and peak surface brightness are $\pm$0.33 per cent and 86~mJy~beam$^{-1}$, respectively. Beam size is $0.34 \times 0.21$ arcsec$^2$ at PA $83\overset{\circ}{.}2$.}
\label{fig:vla-15GHZ-stokesi}
}
\end{figure}

Three distinct sets of radio structures were observed on different spatial scales in NGC\,2329: (1) A one-sided core-jet structure at sub-arcsecond resolutions, (2) a compact core with symmetric, diffuse double-sided inner lobes at arcsecond resolutions, and (3) large lobes consistent with a wide-angle tail radio galaxy on arcminute scales (see Figs \ref{fig:vla-stokesi-collage}, \ref{fig:vla-15GHZ-stokesi}, \ref{fig:vlba-stokesi}, and \ref{fig:combined}).

\subsection{The outer lobes}\label{section:outerlobes}
The 16-arcsec resolution VLA C- and D-array images at {1.5} and {5}~GHz reveal a clear FR~I morphology with two diffuse jets/lobes extending from the core (Fig.~\ref{fig:vla-stokesi-collage}, top panel; we will refer to these as `FR~I' lobes in the rest of this paper). The jets/lobes have almost uniform surface brightness devoid of visible knots or luminosity peaks. The lobes are bent in an obtuse angle, typical for wide-angle tail sources \citep{Owen1976,Smolcic2007}. WATs are preferentially detected in dense, non-relaxed galaxy clusters \citep{Mao2010,Oklopcic2010,Garon2019}. The {north-western} lobe extends upto $\sim$2.9 arcmin~(67~kpc), while the southern lobe has a linear extent of $\sim$2.5 arcmin~(57~kpc). Compared to other FR~I sources, the lobes have smaller lateral dimensions. 

The flux densities of the core in the 16-arcsec image, the north-western lobe, and the southern lobe in the 1.5-GHz VLA C-array and 5-GHz VLA D-array images are listed in Table~\ref{table:vla}; as are the flux densities of the core, the north-western lobe, and the southern lobe in the 4-arcsec resolution 1.5-GHz VLA B-array image (Fig.~\ref{fig:vla-lab-dbcon}, left-hand panel). The flux density estimates in the 1.5-GHz VLA B-array image are lower than those seen in the 1.5-GHz VLA C-array image, due to missing extended emission. To achieve higher sensitivity at higher resolution, we combined the calibrated VLA B- and C-array UV data in \textsc{aips} using the task {\tt DBCON}. We then performed a single round of phase-only self-calibration on the combined data set. The flux densities of the FR~I lobes in the combined VLA B$+$C-array image are noted in Table~\ref{table:vla}; these are consistent with estimates obtained from the 1.5-GHz VLA C-array image.

There is a region of diffuse emission observed by \cite{Feretti1985} at 0.6~GHz between the FR~I lobes. Our higher resolution images do not pick this emission clearly. Two other sources are also visible in this field, which have been identified by \cite{Feretti1985} to be background sources. Using the fractional polarization and fractional polarization noise images, we find that the average fractional polarization of the FR~I lobes is around $30\pm9$ per cent at 1.5~GHz. The average fractional polarization of the `core' which is the Seyfert-like structure, is $16.6\pm2.3$ per cent at 1.5~GHz.

\subsection{The inner lobes}
VLA A- and B-array images at 1.5 and 5 GHz show a core with diffuse bubble-like radio lobes (Fig.~\ref{fig:vla-stokesi-collage}, bottom panel). The lobes have a linear extent of $\sim$5 kpc (15~arcsec) on each side of the core. They appear to have uniform surface brightness. The flux density of the core and lobes at 1.5 and 5 GHz are noted in Table~\ref{table:vla}. The bubble-like radio lobes bear a strong resemblance to those observed in Seyfert galaxies \citep[e.g.][]{Nagar1999,Hota2006,Kharb2006,Croston2008}. These lobes are often suggested to be the result of strong jet--medium interaction {\citep[e.g.][]{Whittle2004,Hota2006,Morganti2015,Oosterloo2017}} which decollimates the jets and even disrupts them completely, thereby explaining why jet-like features are not observed on the 100 pc to kpc-scales in Seyfert lobes \citep{Hota2006,Croston2008,Kharb2014,Sebastian2020}. The double-sided 5-kpc scale lobes resemble the kpc-scale radio emission in Seyfert galaxies \citep[e.g.][]{Baum1993,Gallimore2006,Kharb2006,Singh2015,Congiu2017}. The host galaxy type of S0 for NGC\,2329 is indeed common to Seyfert galaxies. \citet{Kollatschny2008} detected only narrow emission lines in NGC~2329. Based on the observed narrow line ratios, \citet{Kollatschny2008} have classified the host galaxy as a LINER. In the rest of this paper, we will refer to the inner radio lobes as `Seyfert' lobes.

At the intermediate 4-arcsec resolution, the 1.5-GHz VLA image (Fig.~\ref{fig:vla-lab-dbcon}, left-hand panel) resolves both the inner Seyfert-like structure and the outer WAT outflow. The inner Seyfert-like core and the WAT outflows display considerable difference in surface brightness. The difference in surface brightness of the core and the FR~I lobes is further confirmed in the combined high-sensitivity VLA B+C-array image (Fig.~\ref{fig:vla-lab-dbcon}, right-hand panel). The sharp contrast in surface brightness between the inner and outer lobes is consistent with episodic AGN activity in NGC\,2329.
Using the fractional polarization and fractional polarization noise images, we find that the fractional polarization of the entire Seyfert structure at 5~GHz is $15.4\pm4.3$\%, while for the `core' it is $5.1\pm1.2$\%.

\begin{figure*}
\centering{
\includegraphics[width=8.1cm,trim=20 0 0 0]{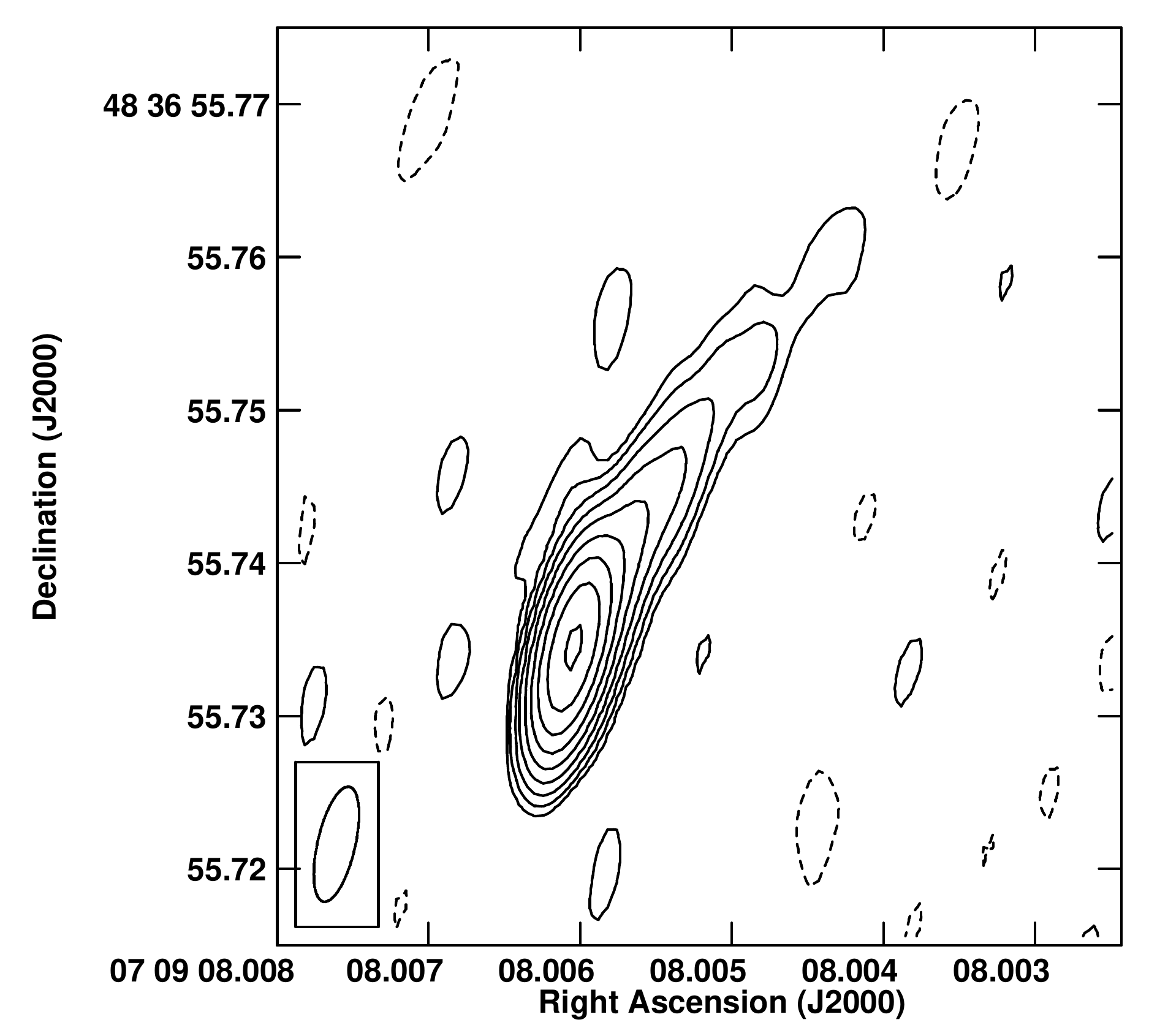}
\includegraphics[width=9.5cm]{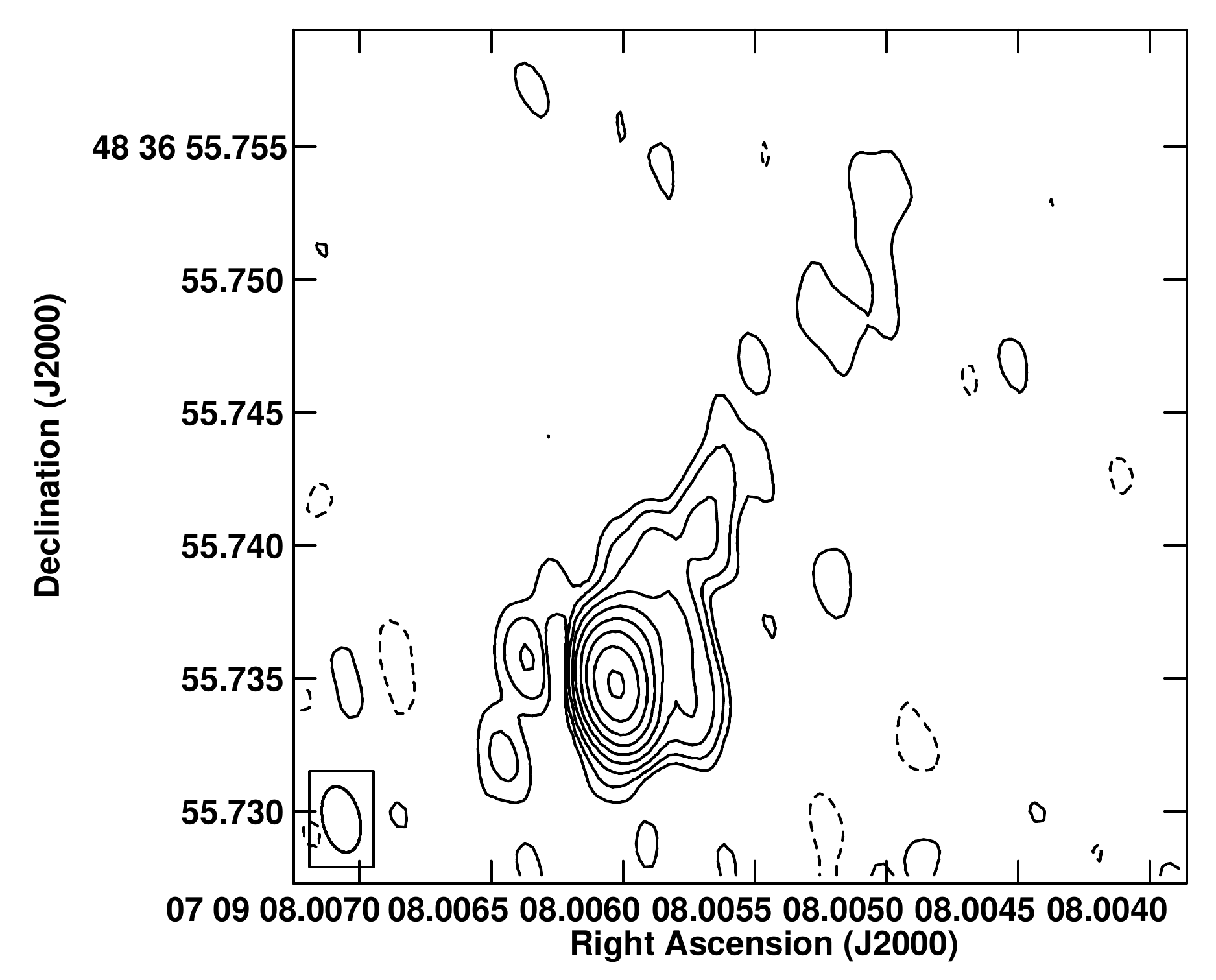}} 
\caption{Left-hand panel: 2.3-GHz VLBA radio contour image of NGC\,2329. Contours are in percentage of peak surface brightness and increase in steps of two. The lowest contour and peak surface brightness are $\pm$0.35 per cent and 50~mJy~beam$^{-1}$. The beam is of size  7.7~mas~$\times$~2.5~mas at PA $-12\overset{\circ}{.}9$. Right-hand panel: 5-GHz VLBA radio contour image. Contours are in percentage of peak surface brightness and increase in steps of two. The lowest contour and peak surface brightness are $\pm$0.35 per cent and 53~mJy beam$^{-1}$. The beam is of size 2.5~mas~$\times$~1.4~mas at PA $12\overset{\circ}{.}7$.}
\label{fig:vlba-stokesi}
\end{figure*}

\begin{figure*}
\centering{
\begin{subfigure}{.50\textwidth}
 \centering
 \includegraphics[width=.99\linewidth]{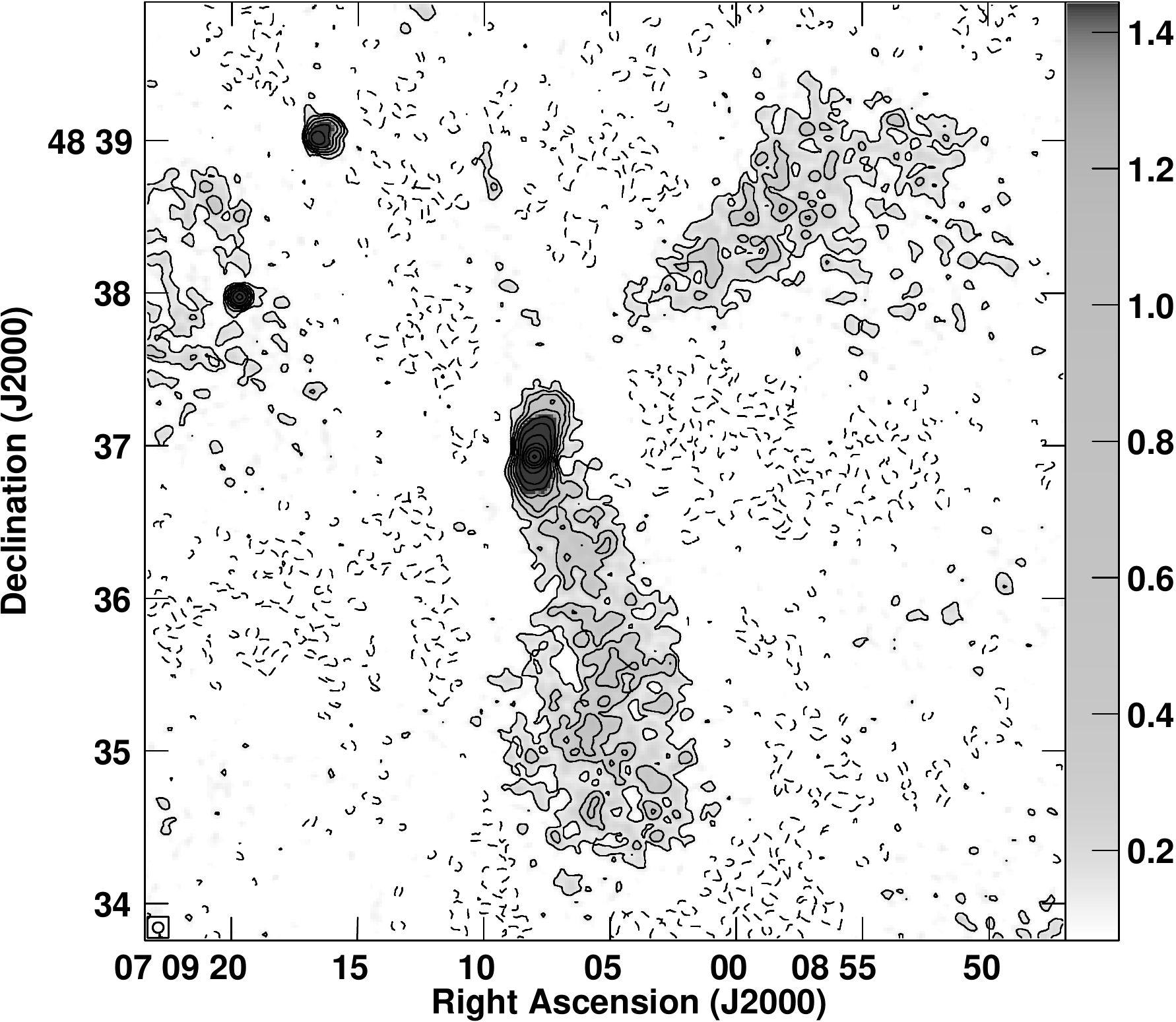}
 \label{subfig:l-b}
\end{subfigure}%
\begin{subfigure}{.50\textwidth}
  \centering
  \includegraphics[trim=1.5cm 1.1cm 1.5cm 1.5cm,clip,width=0.98\linewidth]{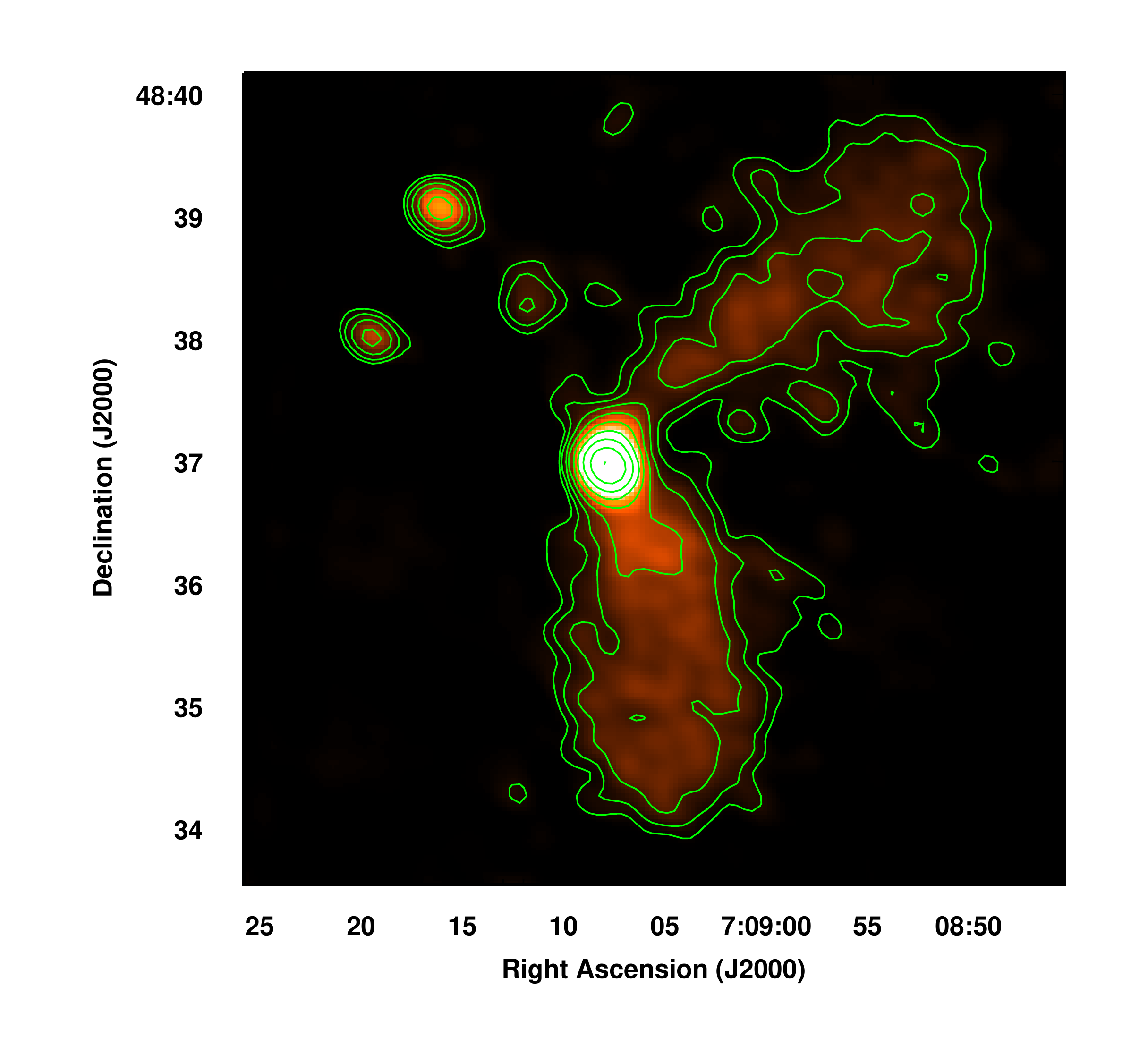}
  \label{subfig:l-bc}
\end{subfigure}}
\caption{
Left-hand panel: 1.5-GHz VLA B-array configuration total intensity image of NGC\,2329. Contours in percentage of peak surface brightness and increasing in steps of two are overlaid on the grey-scale image. The lowest contour and peak surface brightness are $\pm$0.09 per cent and 145~mJy beam$^{-1}$. Beam size is $4.30 \times 4.13$ arcsec$^2$ at PA $-76\overset{\circ}{.}7$. Right-hand panel: high sensitivity 1.5-GHz VLA B$+$C array color image at 16-arcsec resolution overlaid with 1.5-GHz VLA C-array contours. Peak surface brightness of this image is 243~mJy beam$^{-1}$. Beam size is $16.74 \times 13.61$ arcsec$^2$ at PA $60\overset{\circ}{.}8$. Contours are in percentage of peak surface brightness of the VLA C-array image ($=$232~mJy~beam$^{-1}$) and increase in steps of two with the lowest contour level being at $\pm$0.35 per cent.} 
\label{fig:vla-lab-dbcon}
\end{figure*}

\subsection{Sub-kpc and pc-scale jet}
VLBA images at 2.3 and 5~GHz show a bright core with a one-sided core-jet structure indicative of relativistic Doppler boosting in the jet (Fig.~\ref{fig:vlba-stokesi}). The bright core and one-sided jet is consistent with what is observed in FR~I radio galaxy jets \citep{Giovannini1994,Venturi1994,Giovannini2001,Kadler2012,Kharb2012}. The 2.3-GHz core-jet structure extends up to $\sim$12~parsec (0.034~arcsec) in the {north-western} direction, and the 5-GHz structure extends up to $\sim$10 parsec (0.027~arcsec). The pc-scale jet is closely aligned with the optical jet direction at PA $\sim320\degr$. The pc-scale jet aligns well with the WAT lobes. The VLA image at 15~GHz (Fig.~\ref{fig:vla-15GHZ-stokesi}) reveals an $\sim420$~pc (1.1~arcsec) core-jet structure in the north-west direction. The blob-like feature observed towards the west appears to be an image artefact.

We estimated the pc-scale jet speed using the jet-to-counterjet surface brightness ratio ($R_\mathrm{J}$; \cite{Ghisellini1993,Kharb2019}) at 2.3\,GHz. In the absence of a visible counterjet, we used the $3\sigma$  noise level along the counterjet (southern) direction as the limiting counterjet brightness, and subsequently obtained a {lower} limit to the ratio, $R_\mathrm{J}>13.58$. We assumed smooth continuous jet structure with intrinsic {spectral index $=-0.5$ \citep{Giovannini2001}}, and a jet structural parameter $(p) = 3.0$ \citep{Urry1995}. Owing to a lack of broad lines in the optical spectrum \citep{Kollatschny2008}, we considered NGC~2329 to be a Type II object. We assumed that the dusty torus half-opening angle $\sim$50$\degr$ \citep{Simpson1996} and hence the jet inclination $\geq$ 50$^\circ$. Therefore, a lower limit on the jet speed necessary to produce the observed $R_\mathrm{J}$ value is obtained {as $\sim$$0.75c$.} Similar jet speeds are measured in FR~I jets \citep{Venturi1994,Giovannini2001} as well as some Seyfert jets \citep{Brunthaler2000,Middelberg2004}.

\subsection{Spectral index and equipartition estimates}
\begin{figure*}
\centering{
\includegraphics[width=8.55cm]{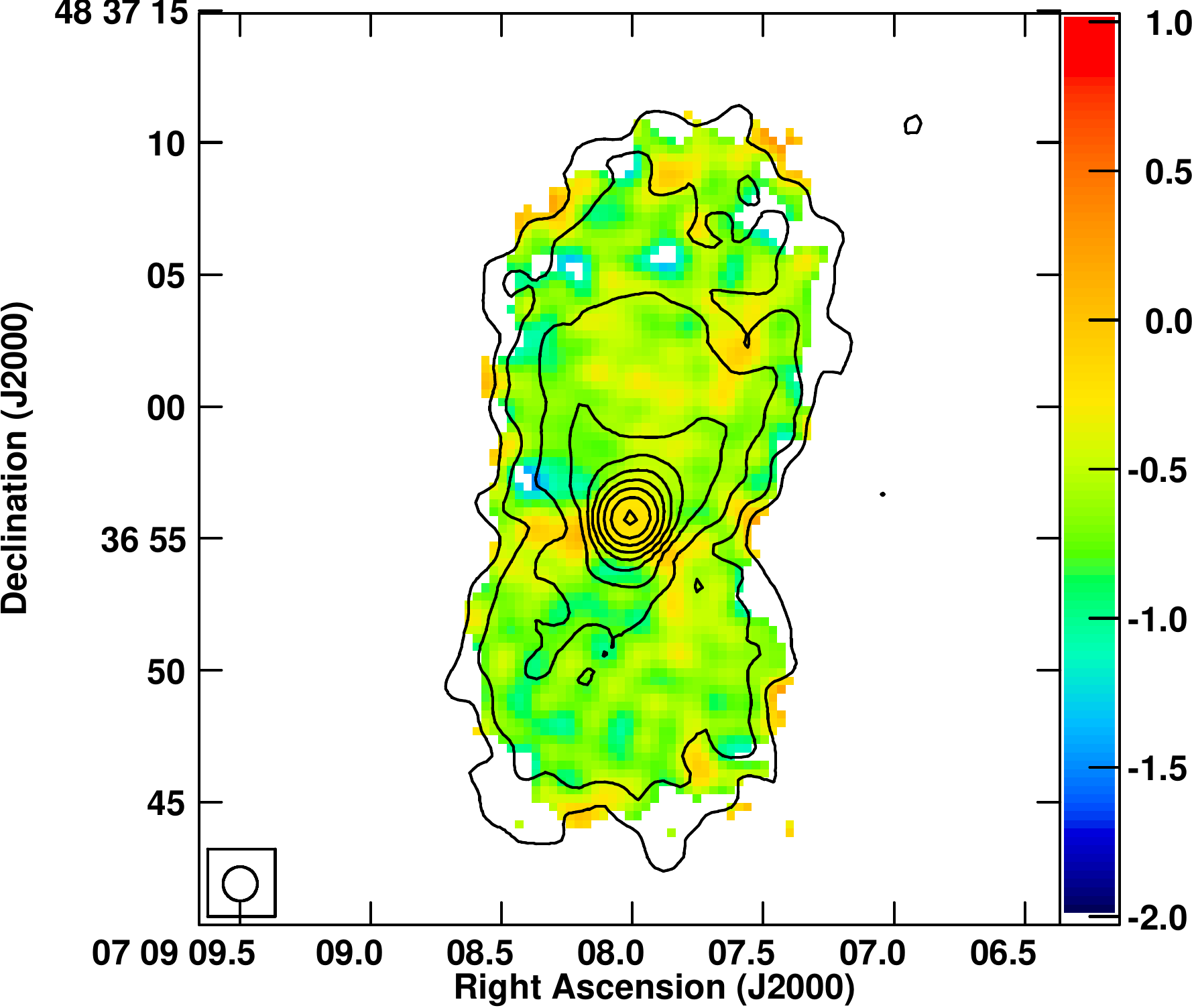}
\includegraphics[width=8.3cm]{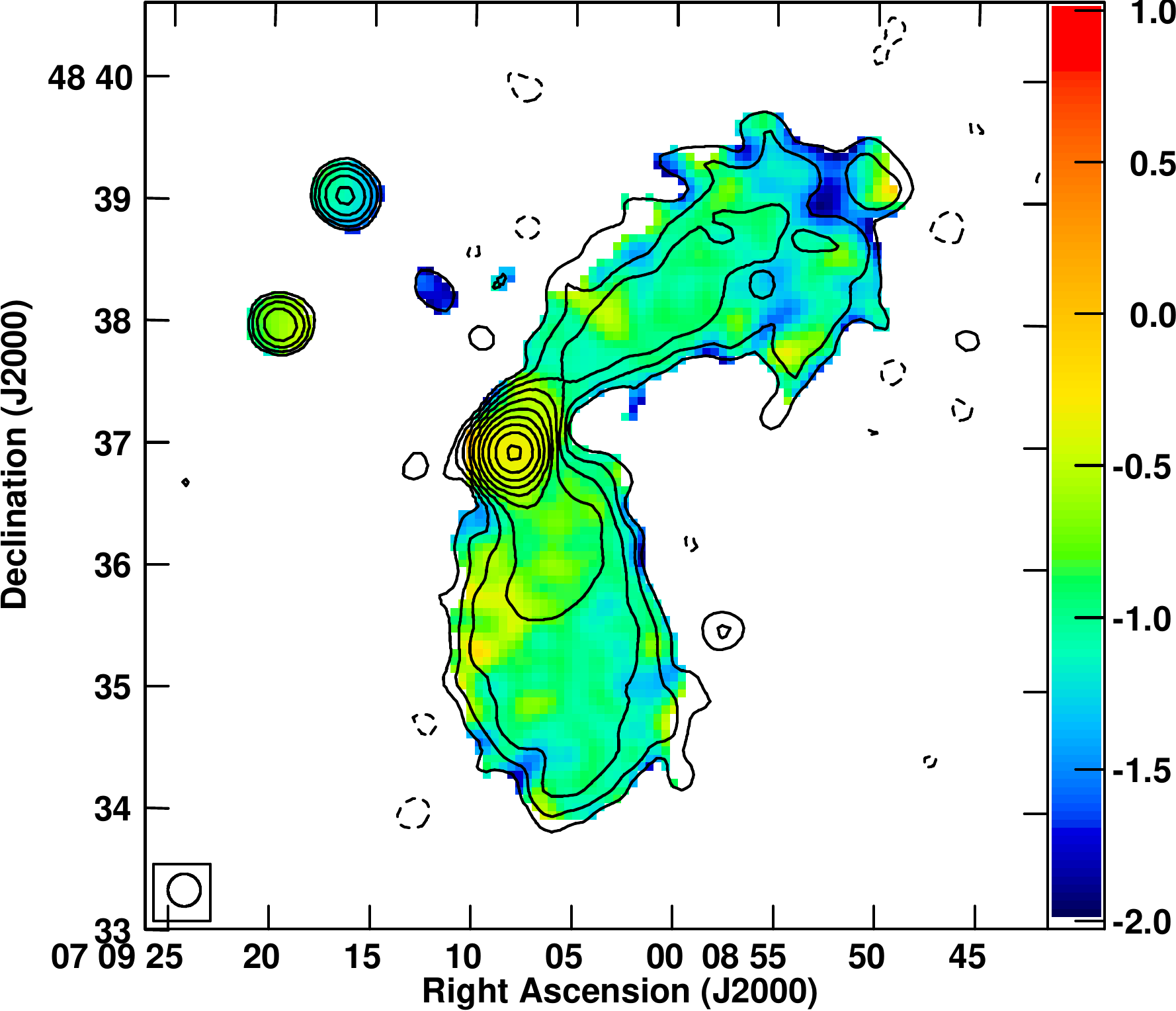}}
\caption{Left-hand panel: the 1-arcsec resolution {1.5--5~GHz} spectral index image in color overlaid with {1.5-GHz} radio contours for NGC\,2329. Contours are in percentage of peak surface brightness and increase in steps of two with the lowest contours being $\pm$0.17 per cent. Right-hand panel: the 16-arcsec resolution {1.5--5~GHz} spectral index image in color overlaid with {5-GHz} radio contours. Contours are in percentage of peak surface brightness and increase in steps of two with the lowest contours being $\pm$0.17 per cent.}
\label{fig:spectind-collage}
\end{figure*}

We computed the spectral index of the 1-arcsec scale Seyfert lobes and the large-scale 16-arcsec FR~I lobes using 1.5- and 5-GHz VLA total intensity images. The mean spectral index {$\alpha^{1.5}_{5}$ of the compact core of the 1-arcsec inner structure is $-0.34\pm0.04$, and that of the diffuse northern and southern Seyfert lobes are $-0.62\pm0.14$ and $-0.65\pm0.16$} respectively. Spectral index remains almost uniform throughout the structure (Fig.~\ref{fig:spectind-collage}, left-hand panel). The relatively flat spectral index of the Seyfert lobes indicate that they are currently powered by the AGN. The FR~I lobes have a significantly steeper spectral index as compared to the Seyfert lobes. The central region of the FR~I structure (Fig.~\ref{fig:spectind-collage}, right-hand panel) has mean spectral index $\alpha^{1.5}_{5}= -0.42\pm0.05$. Mean spectral indices of the {north-western} and southern outflows are $-1.11\pm0.18$, and $-0.96\pm0.13$, respectively. The steeper spectra of the outflows is indicative of an older region, likely to be the remnant of previous episodic activity.

\begin{table*}
\caption{Observed and estimated parameters for the lobes of NGC\,2329.}
\centering
\begin{tabular}{lcccccccc}\hline
& \multicolumn{3}{c}{Observed values}  & \multicolumn{4}{c}{Equipartition parameters} & \multicolumn{1}{c}{Age estimates} \\
& {$S_\mathrm{1.5~GHz}$} & {$S_\mathrm{5~GHz}$} & $\alpha^{1.5}_{5}$ & {$L_\mathrm{rad}$}  & \hspace{0.05in}  {$B_\mathrm{min}$} &  {$E_\mathrm{min}/vol$} &   {$P_\mathrm{min}$} &  \hspace{0.05in}  
$t_{e}$   \\
& {(mJy)} &  {(mJy)} &        &  {($10^{40} \mathrm{erg~s}^{-1}$)}       &       {($\mu G$)} &  {($10^{-13} \mathrm{erg~cm}^{-3}$)} & {($10^{-13} \mathrm{dyn~cm}^{-2}$)} &  {(Myr)}        
\\
& {(a)}& {(b)}&  {(c)} &    {(d)}       & {(e)} &  {(f)} & {(g)} & {(h)} \\ \hline
N.W. FR~I Lobe   & 198 $\pm$ 0.28  & 54 $\pm$ 0.08 & $-1.11$ $\pm$ 0.18 & 1.20 $\pm$ 0.35 & 2.11 $\pm$ 0.24 & 7.11 $\pm$ 1.46 &  4.15 $\pm$ 0.96 & 43 $\pm$ 4 \\
S. FR~I Lobe     & 241 $\pm$ 0.28  & 81 $\pm$ 0.08 & $-0.96$ $\pm$ 0.13 & 1.46 $\pm$ 0.24 & 1.82 $\pm$ 0.14 & 5.26 $\pm$ 0.68 & ~3.07 $\pm$ 0.46 & 42 $\pm$ 5 \\
N. Seyfert Lobe &  63 $\pm$ 0.04  & 29 $\pm$ 0.04 & $-0.62$ $\pm$ 0.14 & 0.38 $\pm$ 0.03 & 7.12 $\pm$ 0.38 & 80.68 $\pm$ 7.00 & 47.06 $\pm$ 5.05 & 25 $\pm$ 4 \\
S. Seyfert Lobe \hspace{0.2cm} &  43 $\pm$ 0.04  & 19 $\pm$ 0.04 & $-0.65$ $\pm$ 0.16 & 0.25 $\pm$ 0.03 & 6.77 $\pm$ 0.38 & 72.87 $\pm$ 6.85 & 42.51 $\pm$ 4.81 & 26 $\pm$ 6 \\\hline
\end{tabular}
\begin{flushleft}
{{\it Notes.} The first two rows refer to large-scale FR~I lobes, while the last two rows refer to the inner Seyfert like lobes. Columns are as follows: {(a)} total flux density at 1.5 GHz; {(b)} total flux density at 5 GHz; {(c)} spectral index; {(d)} total radio luminosity; {(e)} minimum energy magnetic field strength; {(f)} particle energy density per unit volume; {(g)} minimum energy pressure; {(h)} electron lifetime due to synchrotron losses and IC losses over CMB photons.}
\label{table:alpha}
\end{flushleft}
\end{table*}

\begin{figure*}
\centering{
\includegraphics[width=8.8cm,trim=20 160 0 150]{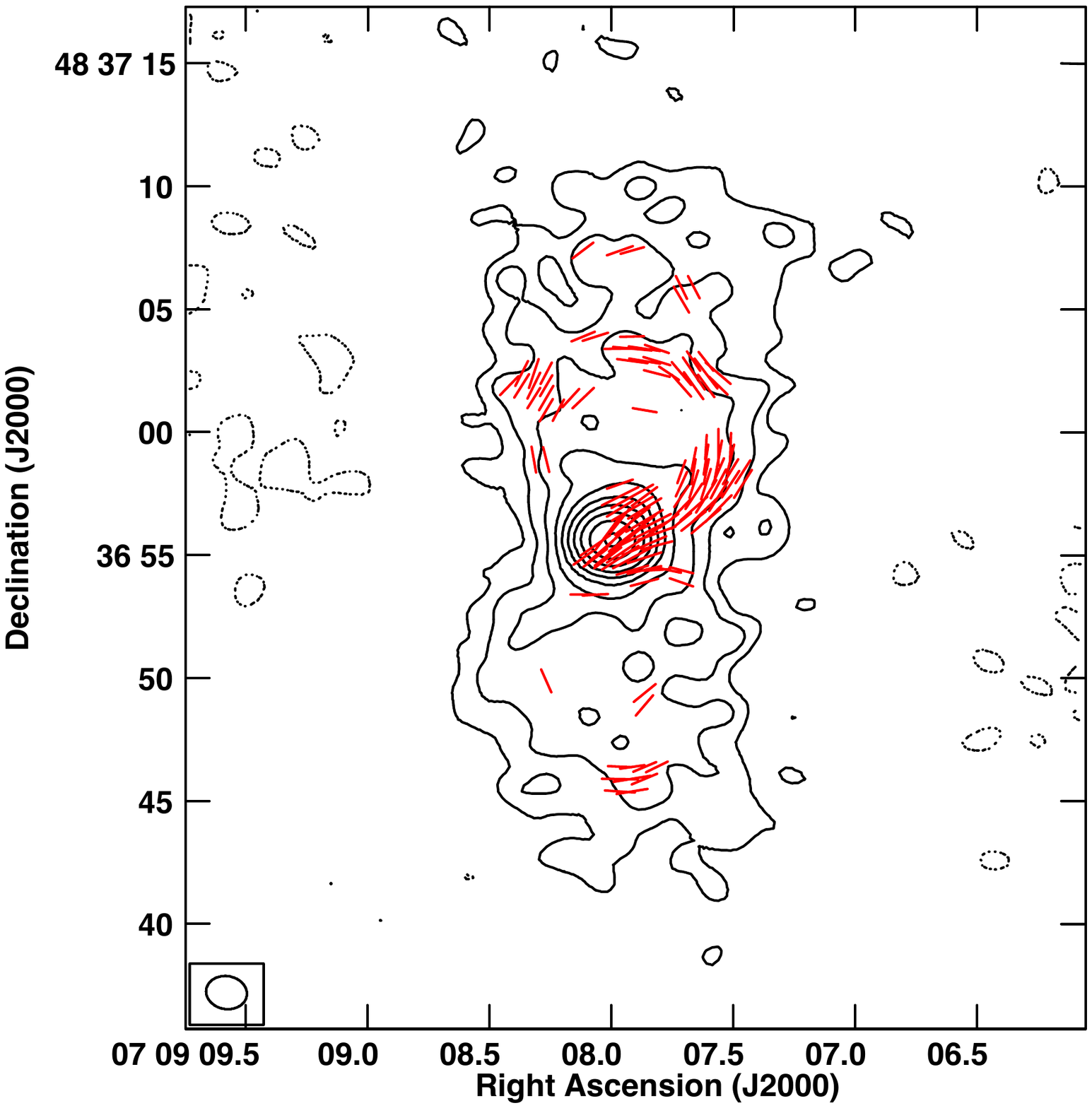}
\includegraphics[width=8.8cm,trim=20 160 0 150]{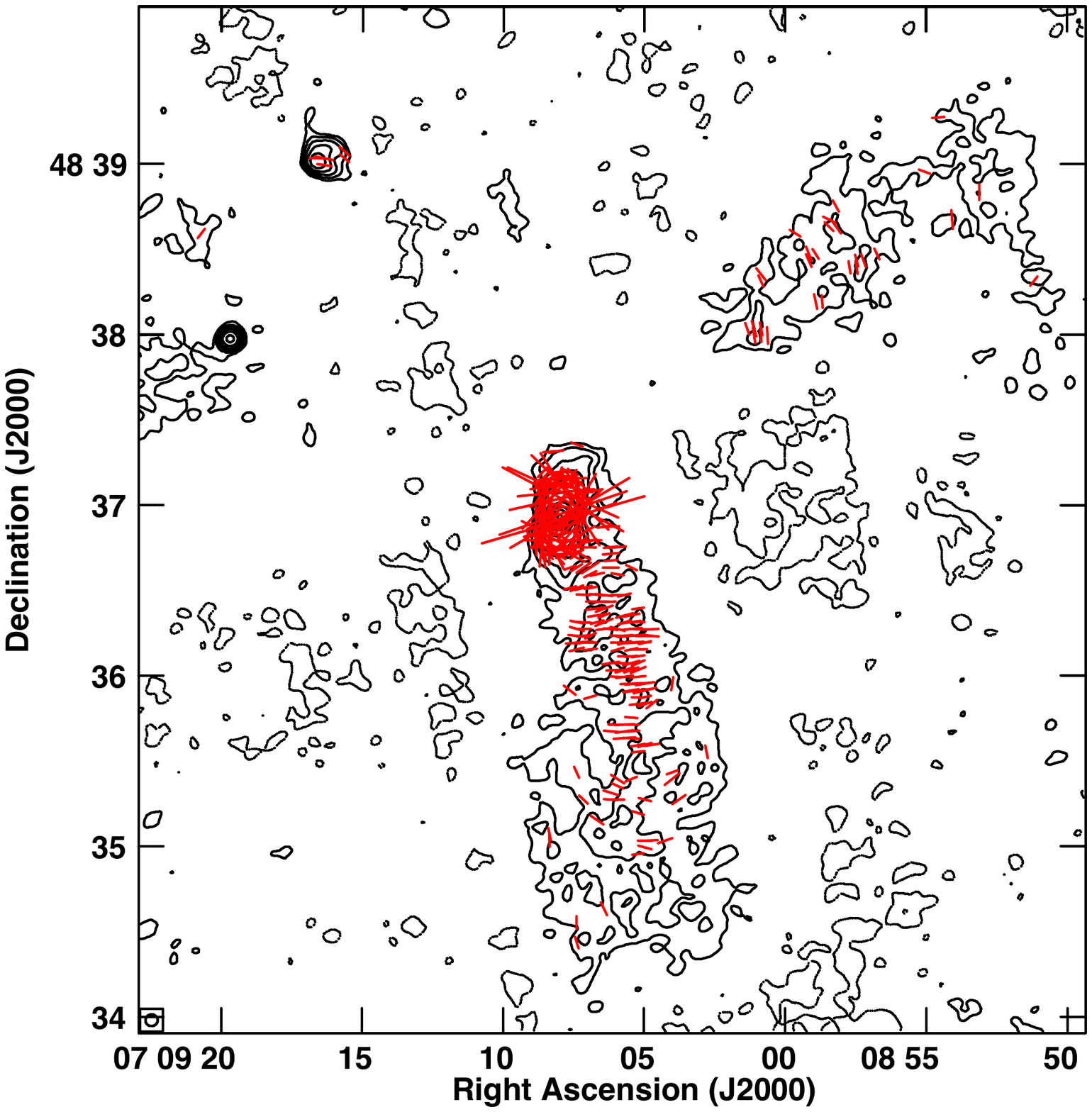}}
\caption{Left-hand panel: VLA B-array 5-GHz radio contours in black with magnetic field vectors superimposed as red ticks whose length is proportional to polarized intensity (length of 2 arcsec corresponds to 0.6~mJy~beam$^{-1}$). Contours are in percentage of peak surface brightness (98~mJy~beam$^{-1}$) and increase in steps of two with the lowest contour being $\pm$0.17 per cent. Right-hand panel: VLA B-array 1.5-GHz radio contours in black with magnetic field vectors superimposed as red ticks whose length is proportional to polarized intensity (length of 20 arcsec corresponds to 1.0~mJy~beam$^{-1}$). Contours are in percentage of peak surface brightness (144~mJy~beam$^{-1}$) and increase in steps of two with the lowest contour being $\pm$0.17 per cent.}
\label{fig:poln}
\end{figure*}

The spectral index across the FR~I lobes in NGC\,2329 is almost uniform without signs of any significant steepening, possibly indicating resupply or re-acceleration by either the AGN or the cluster environment. Using the relations from \cite{Myers1985, ODea1987}, we have calculated the equipartition magnetic field ({$B_\mathrm{min}$}), minimum pressure ({$P_\mathrm{min}$}), and particle energy density ({$E_\mathrm{min}$}) for the central region and the two FR~I lobes of the source. We assumed {the Seyfert lobes and FR~I lobes} to have cylindrical geometry and to be uniformly filled with a volume-filling factor of unity. The relativistic electrons and protons are assumed to have equal energies. By assuming a break frequency of 1.6~GHz, we have calculated the electron lifetime ($t_\mathrm{e}$) due to synchrotron losses due to the equipartition magnetic field ({$B_\mathrm{min}$}) and inverse\,Compton losses off the microwave background with equivalent magnetic field {$B_\mathrm{R}$} \citep{vanderLaan1969}. These parameters {with their corresponding uncertainties} are listed in Table \ref{table:alpha}. The uncertainties arising due to the errors in flux density, spectral indices, and volume of the lobes, were calculated using the standard rules for error propagation \citep[e.g.][equation 3.47]{Taylor1997}.

\subsection{Polarization images and magnetic field structures}
Polarization images at 1.4 and 5 GHz from the WSRT were presented by \citet{Feretti1985}. These indicate the presence of a `spine-sheath' polarization structure in the outer lobes, with inferred magnetic ($B$) fields, assuming optically thin emission, being transverse to the jet/lobe direction in the `spine' and parallel to the jet/lobe direction in the `sheath'. In their 5-GHz polarization image of the inner Seyfert-like structure, the inferred magnetic fields, assuming optically thin emission, appear to be aligned with the lobe edges. In Fig.~\ref{fig:poln}, we have presented the polarization images from the VLA B-array at 1.5 and 5 GHz. Here we have rotated the polarization electric vectors by 90$\degr$ to show the inferred $B$-fields for the optically thin emission. The `spine' of the FR~I jets/lobes with transverse B-fields is clearly visible in the 1.5-GHz image. The `sheath' is not picked up in this high-resolution image. $B$-fields aligned with the Seyfert lobe edges are clearly visible in the 5-GHz image.

\section{Discussion}\label{section:discussion}
Based on differences in surface brightness, radio morphology, spectral indices and polarization of the inner and outer lobes in NGC\,2329, we propose that NGC\,2329 is a restarted AGN. Evidence for multiple epochs of jet activity in radio sources have been reported extensively in the literature \citep{Saikia2009,Brienza2018,Sebastian2019,Sebastian2020,Jurlin2020}. The presence of a strong contrast in surface brightness between the core and lobes, or between the inner and outer pairs of lobes in the case of NGC\,2329, has been taken to indicate restarted activity in radio galaxies, e.g. 4C29.30 \citep{Jamrozy2007}, B2~0258+35 \citep{Shulevski2012}, 3C~293 \citep{Joshi2011}, and 3C~236 \citep{Odea2001}. We have calculated the core prominence (CP) following \citet{Jurlin2020} for NGC\,2329 at 1.5~GHz, $\mathrm{CP_{1.5~GHz}} = S_\mathrm{core}/S_\mathrm{total}$, taking $S_\mathrm{core}$ and $S_\mathrm{total}$ as the flux density of the core and total flux density from the 1.5-GHz VLA C-array image, respectively. $\mathrm{CP_{1.5~GHz}}$ of NGC~2329 comes out to be 0.44, which is consistent with the high $\mathrm{CP_{1.4~GHz}}$ (>0.2) values observed in the restarted sources of \cite{Jurlin2020}.

A large difference is observed in the spectral index of the compact core and extended FR~I-like lobes in the large-scale 16-arcsec structure (Fig.~\ref{fig:spectind-collage}, right-hand panel). The core of {the FR~I structure} has a significantly flatter spectral index $\alpha^{1.5}_{5}= -0.42\pm0.05$ compared to the FR~I lobes, with mean spectral indices $\alpha^{1.5}_{5}$ of the north-western and southern outflows being $-1.11\pm0.18$ and $-0.96\pm0.13$ respectively. This is consistent with restarted radio sources whose newly formed double structure is embedded in the remnant plasma \citep[e.g.][]{Saripalli2002,Jamrozy2007,Murgia2011,Konar2012}. 

Estimates of electron lifetimes ($t_\mathrm{e}$ in Table \ref{table:alpha}) show that the FR~I lobes in NGC\,2329 are about 45-Myr old \citep[consistent with other estimates of ages of FR~I galaxies, see][]{Brienza2017}, while the Seyfert-like lobes are about 25-Myr old. It is evident that the FR~I lobes of NGC\,2329 belong to an older epoch of activity. During this phase, the central AGN restarted its activity cycle and produced a new pair of inner lobes, viz., the smaller Seyfert-like radio lobes seen in the arcsecond scale images. The source seems to have been in a quiescent phase for $\sim$20~Myr. It is worth noting that the inner lobes of NGC\,2329 are not as edge-brightened as observed in some Seyfert galaxies like Mrk\,6 \citep{Kharb2006} or NGC\,6764 \citep{Croston2008}. This may be because the `Seyfert' lobes of NGC\,2329 are expanding within the radio plasma of the previous FR~I lobes. As demonstrated by the numerical simulations of restarted jets by \citet{Clarke1991}, bow shocks (and hotspots) appear fainter or are absent in restarted jets moving into remnant plasma from previous jet episodes. 

\begin{figure}
\centering{
\includegraphics[width=8cm]{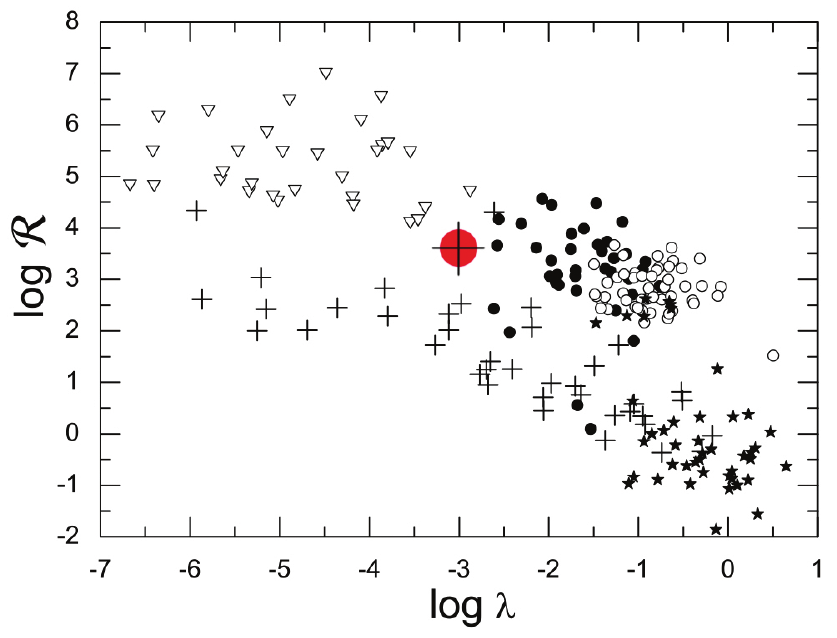}
\caption{ Radio loudness $R$ versus Eddington ratio $\lambda$ plot \citep[fig.~3;][]{Sikora2007}. BLRGs are marked by filled circles, RL quasars by open circles, Seyfert galaxies and LINERs by crosses, FR I radio galaxies by open triangles, and PG quasars by filled stars, respectively, by the original authors. Position of NGC~2329 in this plot has been marked by the red circle and plus symbol.}
\label{fig:sikora}
}
\end{figure}

Numerous reasons have been attributed to explain recurring AGN activity. Accretion disc instability or a change in angular momentum of the accretion disc have been invoked in the past to explain recurrent AGN activity. A change in the jet axis accompanies both disc instability and chaotic accretion \citep{Dennett2002,Hernandez-Garcia2017}. The difference in jet directions of the inner and outer outflows indeed favors such an explanation. However, the direction of the pc-scale VLBI jet makes it clear that the spin axis remains unchanged close to the core. 

\citet{Sikora2007} have argued that radio loudness in an AGN could be proportional to the black hole spin -- the higher the spin, the higher the outflow power, and the higher its radio loudness (also known as the `spin paradigm'). We used the total radio flux density of the large-scale FR~I structure at 5~GHz {(= 344 mJy; Table \ref{table:vla}) and the absolute $B$-band magnitude (= 13.6; Table \ref{table:physicalparam}) to calculate the radio loudness parameter as $R=3855$.} Galaxy Evolution Explorer \citep[{\it GALEX};][]{Seibert2012} source catalogs report near-UV ($1.29\times10^{15}~Hz$) luminosity {($L_\mathrm{NUV}$)}~of~NGC~2329~as~$\mathrm{1.99\times10^{42}~erg~s^{-1}}$. Using {$L_\mathrm{NUV}$}, we obtained the [OIII]\,$\mathrm{\lambda5007}$ luminosity {$L_\mathrm{OIII}$} \citep[equation 5;][]{Stern2012}, thereby calculating bolometric luminosity \citep[$L_\mathrm{bol}/L_\mathrm{OIII}\approx\mathrm{3500}$;][]{Heckman2004} as {$L_\mathrm{bol} = 5.97\times10^{43}$~erg~s$^{-1}$.} We derived the Eddington luminosity ($L_\mathrm{Edd}=\mathrm{6.17\times10^{46}~erg~s^{-1}}$) of this source, and thus obtained the Eddington ratio \citep[$\lambda~=~L_\mathrm{bol}/L_\mathrm{Edd}$;][]{Padovani2017} as $\mathrm{9.68\times10^{-4}}$. 

This places NGC\,2329 at the divide of FR~I and Seyfert galaxies in the $R$ versus $\lambda$ plot of \cite{Sikora2007} (Fig.~\ref{fig:sikora}). In more recent work, \citet{Sikora2013} have invoked the `magnetic flux paradigm' to explain the RL-RQ dichotomy; magnetic flux accumulation that results from a `hot' (radiatively inefficient) accretion disc phase is able to produce large jets in RL AGN. The example of NGC\,2329 suggests that the RL or RQ AGN states could be produced in a single source, given the right accretion disc states. Indeed, RQ Seyfert galaxies, which are expected to be in a soft state corresponding to very high accretion rates, can undergo transitions to low/hard accretion states associated with weak steady radio jets, such as those seen FR~I sources \citep{Meier2012,Blandford2019}. These transitions can occur on time-scales of hundreds to millions of years \citep{Meier2012}.
These accretion disc states could in turn be a function of gas inflow due to minor or major mergers. This suggestion is consistent with the recent works of \citet{Bajraszewska2020,Wolowska2021} who have reported the sudden onset of radio loudness in their sample AGN.

NGC\,2329 displays polarization structures that are consistent with FR~I radio galaxies in its outer lobes and Seyfert galaxies in its inner lobes. `Spine-sheath' polarization structures have been observed in several FR~I radio galaxies on different spatial scales \citep{Kharb2005,Kharb2009,Perlman2006} and could either indicate jet--medium interaction \citep{Laing1996} or the presence of large-scale helical fields \citep{Lyutikov2005}. On the other hand, magnetic fields aligned with the lobe edges have been observed in Seyfert galaxies like Mrk\,6 \citep{Kharb2006}, NGC\,4388 and NGC\,5506 \citep{Sebastian2020}, among others. Therefore, the polarization images clearly underscore the different `nature' of the outflows that make the inner and outer lobes of NGC\,2329. They also suggest a close relationship between Seyfert-like and FR~I-like outflows and perhaps more generally, between RQ and RL AGN.

Strong jet--medium interaction is typically invoked for explaining the morphology of both FR~I radio galaxies and Seyfert galaxies. The interacting medium is typically the IGM in FR~Is and ISM in Seyferts, although interaction with the ISM is also invoked to explain the $\sim$1~kpc-scale jet flaring point observed in several FR~I radio galaxies \citep[e.g.][]{Laing2002}. The core-jet VLBI structure as well as the pc-scale jet speeds in NGC\,2329 appear identical to those observed in regular FR~I radio galaxies \citep{Kharb2012,Kharb2014}. Smaller lateral dimensions of the outflow compared to other FR~I sources is likely due to the presence of dense surrounding medium \citep{Garon2019}. The optical jet-like extension in NGC\,2329 further hints at the presence of a high-pressure environment near the nucleus \citep{Hjorth1995,Macchetto1996,Sparks2000}.

Events such as galaxy mergers \citep{Lara1999,Schoenmakers2000,RamosAlmeida2011}, and cluster mergers \citep{Owen2006,Silverman2008} are potent forces for triggering AGN activity. \cite{Liuzzo2010} found evidence for galaxy merger activity in NGC\,2329. Such a scenario would lead to an infall of gas that would refuel the AGN, subsequently restarting the AGN activity \citep{Acreman2003,Sakelliou2005,Owen2006,Brienza2018}. The exceptional blue $V-I$ color ($\sim$1.20) in NGC\,2329 hints at the presence of a young stellar population. Blue color in bulge-dominated galaxies is directly related to star formation \citep{Silverman2008}. NGC\,2329, however, has a low star formation rate (SFR = $4.11\times10^{-2}~\mathrm{M_{\sun}~yr^{-1}}$) which is consistent with other FR~I radio galaxies \citep{Vaddi2016}. 

Hydrodynamic simulations have shown that shocks driven by radio jets can boost the SFR \citep{Gaibler2012,Silk2013,Mukherjee2018}. Recent works \citep{Fragile2017,Mukherjee2018,Nesvadba2020} have suggested that period of increased star formation triggered by low power jets are short-lived and decline after the short `kick-off' phase. NGC\,2329, which likely restarted with lower jet power, could have seen a similar short-lived increase in star formation, which then slowly declined. This could explain the presence of the young stellar population in spite of a low observed SFR. Another possibility is that the inflow of cool gas fueling star formation activity has been disrupted, leading to a dip in the SFR. Cluster mergers have been identified as potent events that can disrupt cooling flows \citep{Sarazin2002,Chen2017}, affect star formation, and trigger AGN activity \citep{Owen2006}.

Interestingly, the extended FR~I lobes do not show signs of significant spectral steepening. It is possible that the restarted Seyfert lobes may be supplying fresh particles to the ageing FR~I lobes, thereby preventing steepening. Lack of steepening could also be attributed to a combination of mechanisms such as particle re-acceleration, particle mixing, or adiabatic compression. These phenomena might originate due to shocks or stochastic processes, such as turbulence in the inter-cluster medium \citep{Brienza2018}. The lack of brightness enhancement might be caused due to magnetic draping \citep{Adebahr2019}. Magnetic draping intensifies the mixing of differently aged plasma components inside the lobes and enhances diffusive re-acceleration mechanisms \citep{Caprioli2018}. Finally, relatively uniform spectral indices in the radio galaxy NGC\,3998 \citep{Sridhar2019} caused by AGN sputtering could explain similar uniform spectral indices inside the lobes of NGC\,2329. 

\begin{figure*}
\centering
\includegraphics[trim=2.8cm 10.4cm 1.6cm 2.0cm,clip,width=16cm]{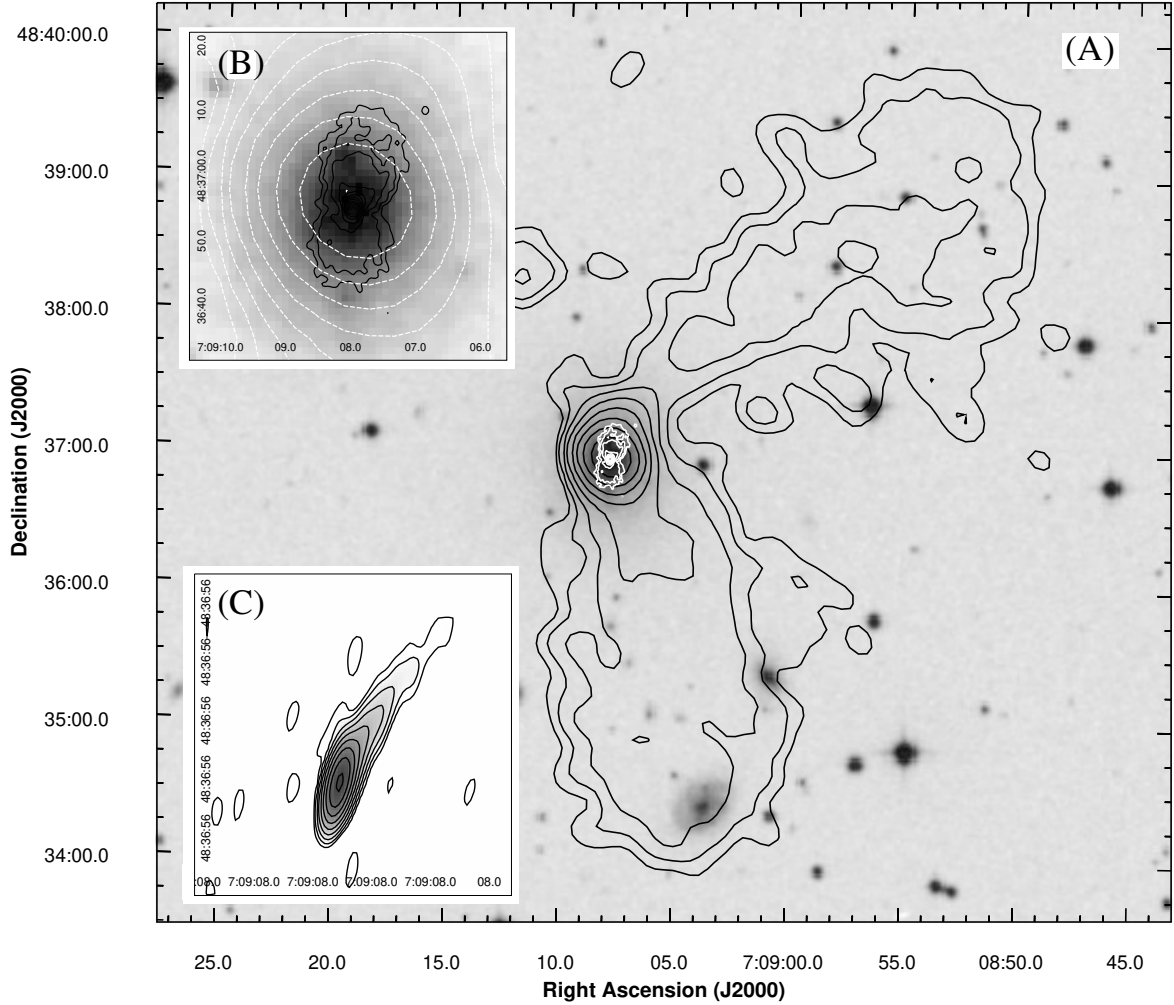}
\caption{Panel (a): VLA radio contour images of NGC\,2329 at different scales overlaid on the DSS POSS2 optical image. 1.5-GHz 16-arcsec resolution VLA contours are denoted by the black lines, and 1.5-GHz 1-arcsec resolution VLA contours by white. Panel (b): zoomed-in snapshot of the central region showing the 1.5-GHz 1-arcsec resolution VLA contours (black) over the 1.5~GHz 16-arcsec VLA contours (white dashed). Panel (c): VLBA 2.3-GHz radio contours overlaid on grey-scale VLBA 2.3-GHz total intensity image.}
\label{fig:combined}
\end{figure*}

\section{Conclusions}\label{section:conclusion}
In this paper, we have presented the radio images of the wide-angle tailed galaxy NGC\,2329 at multiple frequencies and resolutions, using archival data from the VLA and VLBA. We have found that the source exhibits a composite radio morphology -- a Seyfert-like double-bubble structure on arcsecond scales, and a wide-angle tailed FR~I radio galaxy on arcminute scales. The FR~I lobes appear significantly dimmer than the central structure. Spectral index images of both the structures show nearly uniform spectral indices without signs of significant variations inside them. The central structure has a relatively flat spectrum, while the FR~I lobes have a steeper spectrum, implying their origin from two different AGN activity episodes.

It is interesting to note that NGC\,2329 exhibits other intermediate properties between FR~Is and Seyferts. It lies in between the FR~I and Seyfert zones in the radio loudness vs Eddington ratio (or accretion rate) plot of \citep{Sikora2007}, and exhibits a pc-scale jet speed that is intermediate as well. {Most interestingly, the polarization and inferred magnetic field structures observed in the outer and inner lobes resemble those observed in bona fide FR~I radio galaxies (`spine-sheath' B-fields) and Seyfert galaxies (B-fields aligned with the lobes edges), respectively.} The nearly uniform spectral indices observed in the radio lobes of both the Seyfert and FR episodes is consistent with AGN sputtering as observed in NGC\,3998 by \citet{Sridhar2019}. 

We have calculated the `minimum energy' parameters pertaining to the central structure as well as the two FR~I lobes. Estimates of the characteristic electron lifetimes, suffering radiative and adiabatic losses, suggest that the FR~I lobes are at least twice as old as the Seyfert lobes. This suggests that the AGN initially created an FR~I-like jet, became inactive, and then restarted, producing a Seyfert-like outflow. 

We suggest the cluster merger in Abell 569 and strong jet--medium interaction in NGC\,2329 to be primarily responsible for the recurrent AGN activity as well as the curious composite radio structure observed in NGC\,2329. The first AGN activity episode resulted in the FR~I WAT radio structure. The AGN activity stopped eventually. In the intervening quiescent period of AGN activity, an inflow of copious amounts of gas occurred and provided the accretion fuel. This resulted in a new AGN activity episode as well as greater jet frustration for the new radio outflow leading to a Seyfert-like structure. These merger-related gas inflows may have influenced the accretion disc states that may be `hot' (radiatively inefficient) during the FR~I phase and turned into `cold' (radiatively efficient) state during the Seyfert phase \citep[e.g.][]{Meier2012,Sikora2013,Blandford2019}. NGC\,2329 therefore supports the idea that the RL/RQ dichotomy could be a function of time in the lifecycle of an AGN.

\section*{Acknowledgments}
We thank the referee for insightful suggestions that have improved the manuscript significantly.
SD is thankful to NCRA for a position in its Visiting Students' Research Programme (VSRP), 2018. PK acknowledges the Helena Kluyver fellowship that made the  collaborative visit to ASTRON possible, greatly propelling this project. SN acknowledges support by the Science \& Engineering Research Board, a statutory body of Department of Science \& Technology (DST), Government of India (FILE NO.PDF/2018/002833). The National Radio Astronomy Observatory is a facility of the National Science Foundation operated under cooperative agreement by Associated Universities, Inc. This work has made use of the NASA/IPAC Extragalactic Database (NED), which is operated by the Jet Propulsion Laboratory, California Institute of Technology, under contract with the National Aeronautics and Space Administration. We thank the staff of the GMRT that made these observations possible. GMRT is run by the National Centre for Radio Astrophysics of the Tata Institute of Fundamental Research. We acknowledge the support of the Department of Atomic Energy, Government of India, under the project 12-R\&D-TFR-5.02-0700.

\section*{Data Availability}
VLA data sets (Project IDs AB0920, AC0445, AB0412, AC0557, and AS0451) and VLBA data sets (Project IDs BS0103 and BO0015) used for the analysis and findings within this paper were derived from \href{https://science.nrao.edu/facilities/vla/archive/index}{NRAO VLA Data Archive}, and the DSS optical image was obtained from \href{https://archive.stsci.edu/cgi-bin/dss_form}{The STScI Digitized Sky Survey}. Derived data and images shall be shared by the corresponding author upon reasonable request.

\bibliographystyle{mnras}
\bibliography{sdas2020_ngc2329} 

% Don't change these lines
\bsp    % typesetting comment
\label{lastpage}
\end{document}